\documentclass[12pt]{article}
\usepackage{graphicx}
\usepackage{epstopdf}
\usepackage{subfigure}
\usepackage{amsmath}
\usepackage{amssymb}
\usepackage{amsthm}
\usepackage{enumerate}
\usepackage{booktabs}

\begin{document}

\title{Superconducting properties of 3D low-density TI-bipolaron gas in magnetic field}

\author
{Victor D. Lakhno,$^{1\ast}$\\
\\
\normalsize{$^{1}$Keldysh Institute of Applied Mathematics }\\
\normalsize{of Russian Academy of Sciences, 125047, Moscow, Russia}\\
\\
\normalsize{$^\ast$To whom correspondence should be addressed; E-mail:  lak@impb.ru}
}

\maketitle

\begin{abstract}
Consideration is given to thermodynamical properties of a three-dimensional Bose-condensate of translation-invariant bipolarons (TI-bipolarons) in magnetic field.  The critical temperature of transition, critical magnetic fields, energy, heat capacity and the transition heat of TI-bipolaron gas are calculated. Such values as maximum magnetic field, London penetration depth and their temperature dependencies are calculated. The results obtained are used to explain experiments on high-temperature superconductors.
\end{abstract}

\textbf{Keywords:} Bose-Einstein condensation (BEC); Bardeen-Cooper-Schrieffer theory (BCS); translation-invariant bipolarons; electron-phonon interaction; energy gap

\section{Introduction}\label{Introduction}

Before the discovery of high-temperature superconductivity (HTSC) Bardeen-Cooper-Schrieffer theory~\cite{1} (BCS) played the role of fundamental microscopic theory of superconductivity with, in fact, no alternative. The discovery of HTSC revealed some problems which arose while trying to describe various properties of high-temperature superconductors within BCS. This gave birth to a great number of alternative theories aimed at resolving the problems. Review of various HTSC theories is presented in numerous papers. The current state-of-the-art can be found in~\cite{2,3,4,5,6,7,8,9,10,11,12,13}. All the approaches, are however, based on the same proposition the phenomenon of bosonization of Fermi-particles, or Cooper effect. This proposition straightforwardly leads to the conclusion that the phenomenon of superconductivity is related to the phenomenon of Bose-Einstein condensation (BEC). Presently the idea that superconductivity is based on BEC is generally recognized.

A great obstacle to the development of the theory which should be based on BEC was a statement made in BCS
(see comment in \cite{1}, p. 1177 and~\cite{14}) about incompatibility of their theory with BEC.

Some evidence that this viewpoint is erroneous was first obtained in paper~\cite{15} whose authors, while studying the properties of high-density exciton gas, demonstrated an analogy between BCS theory and BEC. The results of~\cite{15} provided the basis for developing the idea of crossover passing on from the BCS theory which is appropriate for the limit of weak electron-phonon interaction (EPI) to BEC which is suitable for the limit of strong EPI~\cite{16,17,18,19,20,21,22}. It was believed that additional evidence in favor of this approach is Eliashberg theory of strong coupling~\cite{23}. According to~\cite{24}, in the limit of infinitely strong EPI this theory leads to the regime of local pairs, though greatly different from the regime of BEC~\cite{25}.

The attempts to develop a theory of crossover between BCS and BEC ran, however, into insurmountable obstacles. Thus, for example, an idea was put forward, to construct a theory with the use of T-matrix transition where T-matrix of the initial Fermion system would transform into T-matrix of Boson system as the force of EPI would increase \cite{26,27,28,29,30,31}. The approach, however, turned out to be unfeasible even in the case of greatly diluted systems. Actually, the point is that when the system consists of only two fermions it is impossible to construct a one-boson state of them. In the EPI theory this problem is known as that of a bipolaron.

The reason the crossover theory failed can be the following: The BCS theory, as the bipolaron theory, proceeds from Froehlich Hamiltonian. For this Hamiltonian, an important theorem of analyticity of the polaron and bipolaron energy on EPI constant
is proved~\cite{32}. In the BCS theory a very important approximation is, however, made. Namely the actual matrix element of the interaction in Froehlich Hamiltonian is replaced by a model quantity a matrix element truncated near Fermi surface. This procedure is by no means fair. As is shown in~\cite{33}, in the bipolaron theory it leads to ghost effects existence of a local energy level separated by a gap from the quasicontinuous spectrum (Cooper effect). This solution is isolated and does not possess the property of analyticity on coupling constant. In the BCS theory just this solution provides the basis for constructing the superconductivity theory.

As a result, the theory constructed and its analytic continuation (Eliashberg theory) greatly falsify the reality, in particular, they make impossible development of the superconductivity theory on the basis of BEC. Replacement of the actual matrix element by the model one enables one to make analytical calculations completely. Thus, substitution of local interaction for actual one in BCS enabled the authors of~\cite{34} to derive Ginzburg-Landau (GL) phenomenological model which is also a local model. Actually, the power of this approach can hardly be overestimated since it has enabled one to get a lot of statements consistent with the experiment.

This paper is an attempt to develop a HTSC theory on the basis of BEC of translation-invariant bipolarons (TI-bipolarons)~\cite{33,35,36,37,38}, free of approximations made in~\cite{1}.

If we proceed from the fact that the superconducting mechanism is based on Cooper pairing, then in the case of a strong Froehlich electron-phonon interaction this leads to translation-invariant bipolaron theory of HTSC~\cite{Lakhno-ACMP,lakhno-physC}. In distinction from  bipolarons with broken symmetry, TI-bipolaron is delocalized in space and polarization potential well is lacking (zero polarization charge). The energy of TI-bipolarons is lower than the energy of bipolarons with broken symmetry, so they are more stable than bipolarons with broken symmetry.

We recall the main results of the theory of TI-polarons and bipolarons obtained in~\cite{33,35,36,37,38}. Notice that consideration of just electron-phonon interaction is not essential for the theory and can be generalized to any type of interaction, for example the interaction of electrons with the spin subsystem~\cite{39}.

In what follows, we will deal only with the main points of the theory important for the HTSC theory.
The main result of papers~\cite{33,35,36,37,38} is the construction of delocalized polaron and bipolaron states in the limit of strong electron-phonon interaction. The theory of TI-bipolarons is based on the theory of TI-polarons~\cite{40} and retains the validity of basic statements proved for TI-polarons. The~chief of them is the theorem of analytic properties of the ground state of a TI-polaron (accordingly TI-bipolaron) depending on the constant of electron-phonon interaction $\alpha$.  The main implication of this statement is the absence of a critical value of the EPI constant $\alpha _c$ below which the bipolaron state becomes impossible since it decays into independent polaron states. In other words, if there exists a value of $\alpha _c$ at which the TI-state becomes energetically disadvantageous with respect to its decay into individual polarons, then nothing occurs at this point but for $\alpha < \alpha _c$ the state becomes metastable. For~this reason we can expect that for $\alpha < \alpha _c$ the normal phase occurs rather than the superconducting~one.

Another important property of TI-bipolarons is the possibility of changing the correlation length over the whole range of $[0,\infty]$ variation depending on the Hamiltonian parameters~\cite{36,38}. Hence, it can be both much larger (as is the case in metals) and much less than the characteristic size between the electrons in an electron gas (as happens with ceramics).

An outstandingly important property of TI-polarons and bipolarons is the availability of an energy gap between their ground and excited states (Section~\ref{Energy}).

The above-indicated characteristics can be used to develop a microscopic HTSC theory on the basis of TI-bipolarons.

The paper is arranged as follows. In Section~\ref{Pekar} we take Pekar-Froehlich Hamiltonian for a bipolaron as an initial Hamiltonian. The results of three canonical transformations, such as Heisenberg transformation, Lee-Low-Pines transformation and that of Bogolyubov-Tyablikov are briefly outlined. Equations determining the TI-bipolaron spectrum are derived.
In Section~\ref{Energy} we analyze solutions of the equations for the TI-bipolaron spectrum. It is shown that the spectrum has a gap separating the ground state of a TI-bipolaron from its excited states which form a quasicontinuous spectrum. The concept of an ideal gas of TI-bipolarons is substantiated.

With the use of the spectrum obtained, in Section~\ref{Stat} we reproduce thermodynamic characteristics of an ideal gas of TI-bipolarons in the absence of a magnetic field, considered earlier in~\cite{Lakhno-ACMP,lakhno-physC}.

In Section~\ref{Current} we deal with the case when the external magnetic field differs from zero. It is shown that the current state in the system under discussion is caused by the existence of a constant quantity the total momentum of the electron-phonon system in the magnetic field. Comparison of the value of the total momentum with that obtained within phenomenological approach enables us to determine the London penetration depth which is a very important characteristic. The results of the initial isotropic model are generalized to anisotropic case.

In Section~\ref{Thermodyn} we investigate thermodynamic characteristics of an ideal TI-bipolaron gas in the presence of a magnetic field. It is shown that the availability of an energy gap in the TI-bipolaron spectrum makes possible their  Bose-condensation in a magnetic field. A notion of a maximum value of the magnetic field intensity for which homogeneous Bose-condensation is possible is introduced. The temperature dependence of the value of the critical magnetic field and the dependence of the critical temperature on the magnetic field are found. It is shown that the phase transition of an ideal TI-bipolaron gas can be either of the 1-st kind or of infinite kind, depending on the magnetic field value. The theory is generalized to the case of anisotropic superconductor. This generalization enables us to compare the results obtained with experimental data (Section~\ref{Comparison}).

In Section~\ref{Scaling} we consider scaling relations in superconductors. Alexandrov's formula and Homes's law are derived.

In Section~\ref{Summary} the results obtained are summed up.

\section{Pekar-Froehlich Hamiltonian. Canonical Transformations}\label{Pekar}

Following~\cite{33,35,36,37}, in describing bipolarons we will proceed from Pekar-Froehlich Hamiltonian with non zero magnetic field:

\begin{eqnarray}\label{1}
    H=\frac{1}{2m^*}{\left(\hat{p}_1-\frac{e}{c}\vec{A}_1\right)^2}+\frac{1}{2m^*}{\left(\hat{p}_2-\frac{e}{c}\vec{A}_2\right)^2}+
		\sum_k{\hbar\omega_0(k) a^+_k a_k}+\\ \nonumber
		\sum_k\left(V_k e^{i\vec{k}\vec{r}_1}a_k+V_k e^{i\vec{k}\vec{r}_2}a_k+H.c.\right)+
		U\left(\left|\vec{r}_1-\vec{r}_2\right|\right),\\ \nonumber
		U\left(\left|\vec{r}_1-\vec{r}_2\right|\right)=
		\frac{e^2}{\epsilon_{\infty}\left|\vec{r}_1-\vec{r}_2\right|},
\end{eqnarray}
where $\hat{p}_1,\vec{r}_1$, $\hat{p}_2,\vec{r}_2$ are momenta and coordinates of the
first and second electrons, respectively;
$a^+_k$, $a_k$ are operators of the creation and annihilation of the field quanta with energy $\hbar\omega_0(k)$;
$m^*$ is the electron effective mass; the quantity $U$ describes Coulomb repulsion between the electrons;
$V_k$ is the function of the wave vector $k$. We write \eqref{1} in general form. For a special case of ion crystals which is typical for HTSC we consider the Pekar-Froehlich Hamiltonian in the form:

\begin{eqnarray}\label{2}
     V_k=\frac{e}{k}\sqrt{\frac{2\pi\hbar\omega_0}{\tilde{\epsilon}V}}=
		\frac{\hbar\omega_0}{ku^{1/2}}\left(\frac{4\pi\alpha}{V}\right)^{1/2},\ \
		u=\left(\frac{2m^*\omega_0}{\hbar}\right)^{1/2},\ \
		\alpha=\frac{1}{2}\frac{e^2u}{\hbar\omega_0\tilde{\epsilon}},\\ \nonumber
		\tilde{\epsilon}^{-1}=\epsilon^{-1}_{\infty}-\epsilon^{-1}_{0},
\end{eqnarray}
where $e$ is the electron charge; $\epsilon _{\infty}$ and $\epsilon _0$ are high-frequency and static dielectric permittivities;
$\alpha$ is the constant of electron-phonon interaction;
$V$ is the system's volume, $\omega_0$ is the optical phonon frequency.

The axis $z$ is chosen along the direction of the magnetic field induction $\vec{B}$ and use is made of symmetrical gauge:
$$\vec{A}_j=\frac{1}{2}\vec{B}\times\vec{r}_j,$$
for $j=1.2$ . For the bipolaron singlet state discussed below, the contribution of the spin term is equal to zero.

In the system of the center of mass Hamiltonian \eqref{1} takes the form:
\begin{eqnarray}\label{3}
     H=\frac{1}{2M_e} {\left(\hat{p}_R-\frac{2e}{c}\vec{A}_R\right)^2}+
		\frac{1}{2\mu_e} {\left(\hat{p}_r-\frac{e}{2c}\vec{A}_r\right)^2}+
		\sum_k{\hbar\omega_0(k) a^+_k a_k}+\\ \nonumber
		\sum_k{2V_k \text{cos} \frac{\vec{k}\vec{r}}{2}}\left(a_k e^{i\vec{k}\vec{R}}+H.c.\right)+
		U(|\vec{r}|),
\end{eqnarray}
$$\vec{R}=(\vec{r}_1+\vec{r}_2)/2,\ \ \vec{r}=\vec{r}_1-\vec{r}_2,\ \ M_e=2m^*,\ \ \mu_e=m^*/2,$$
$$\vec{A}_r=\frac{1}{2}B(-y,x,0),\ \ \vec{A}_R=\frac{1}{2}B(-Y,X,0),$$
$$\hat{p}_R=\hat{p}_1+\hat{p}_2=-i\hbar\nabla_{\vec{R}},\ \ \hat{p}_r=(\hat{p}_1-\hat{p}_2)/2=-i\hbar\nabla_{r},$$
where $x$,$y$, and $X$, $Y$ are components of vectors $\vec{r}$,$\vec{R}$ respectively.

Let us subject Hamiltonian $H$ to Heisenberg canonical transformation~\cite{41,42}:
\begin{equation}\label{4}
    S_1=\text{exp}\ i\ \left(\vec{G}-\sum_k\vec{k}a^+_k a_k\right)\vec{R}.
\end{equation}

\begin{equation}\label{5}
    \vec{G}=\hat{\vec{\mathcal{P}}}_R+\frac{2e}{c}\vec{A}_R,\ \ \
		\hat{\vec{\mathcal{P}}}_R=\hat{\vec{p}}_R+\sum_k\hbar\vec{k}a^+_ka_k,
\end{equation}
where $\vec{G}$ is the quantity commuting with the Hamiltonian, thereby being a constant, i.e., $c$-number,
$\hat{\vec{\mathcal{P}}}_R$ is the total momentum.

Action of $S_1$ on the field operator yields:
\begin{equation}\label{6}
    S_1^{-1}a_kS_1=a_ke^{-i\vec{k}\vec{R}},\ \ \ S_1^{-1}a_k^+S_1=a_k^+e^{-i\vec{k}\vec{R}}.
\end{equation}

Accordingly, the transformed Hamiltonian $\tilde{H}=S^{-1}_1HS_1$ takes on the form:
\begin{eqnarray}\label{7}
     \tilde{H}=\frac{1}{2M_e} {\left(\vec{G}-\sum_k\hbar\vec{k}a^+_ka_k-\frac{2e}{c}\vec{A}_R\right)^2}+
		\frac{1}{2\mu_e} {\left(\hat{\vec{p}}_r-\frac{e}{2c}\vec{A}_r\right)^2}+
		\sum_k{\hbar\omega_0(k) a^+_k a_k}+\\ \nonumber
		\sum_k{2V_k \text{cos} \frac{\vec{k}\vec{r}}{2}}\left(a_k +a^+_k\right)+
		U(|\vec{r}|).
\end{eqnarray}

In what follows we will believe:
\begin{equation}\label{8}
   \vec{G}=0.
\end{equation}

The physical meaning of \eqref{8} is the absence of the total momentum (current) in the bulk of superconductor. This fact follows from the Meissner effect which states that the current in the volume of superconductor needs to be zero. We use this fact in Section~\ref{Current} to obtain the value of London penetration depth $\lambda$.
We will seek the solution of the stationary Schroedinger equation corresponding to Hamiltonian \eqref{7} in the form:
\begin{eqnarray}\label{9}
   \Psi_H(r,R,\{a_k\})=\phi(R)\Psi_{H=0}(r,R,\{a_k\})\\
   \phi(R)=\exp\left(-i\frac{2e}{\hbar c}\int^{\vec{R}}_0\vec{A}_{R'}d\vec{R'}\right) \notag \\
   \Psi_{H=0}(r,R,\{a_k\})=\psi(r)\Theta(R,\{a_k\}), \notag
\end{eqnarray}
where $\Psi_{H=0}(r,R,\{a_k\})$ is bipolaron wave function in the absence of magnetic field. The explicit form of functions $\psi(r) \text{ and } \Theta(R,\{a_k\})$ is given in \cite{35,38}.

Averaging of $\tilde{H}$ over the wave function $\phi(R) \text{ and } \psi(r)$ yields:
\begin{eqnarray}\label{10}
     \bar{\tilde{H}}=\frac{1}{2M_e} {\left(\sum_k\hbar\vec{k}a^+_ka_k\right)^2}+
		\sum_k{\hbar\tilde{\omega}_k a^+_k a_k}+
		\sum_k \bar{V}_k  \left(a_k +a^+_k\right)+\bar{T}+\bar{U}+\bar{\Pi},
\end{eqnarray}
where:
\begin{eqnarray}\label{11}
    \bar{T}=\frac{1}{2\mu_e}\left\langle \psi\left| (\hat{p}_r-\frac{e}{2c}\vec{A}_r)^2\right|
		\psi\right\rangle, \ \ \
		\bar{U}=\left\langle \psi\left|U(r)\right|\psi\right\rangle ,\ \ \
		\bar{V}_k=2V_k\left\langle \psi\left|\text{cos}\frac{\vec{k}\vec{r}}{2}
		\right|\psi\right\rangle ,\\ \nonumber
		\bar{\Pi}=\frac{2e^2}{M_ec^2}\left\langle \phi\left|A^2_R\right|\phi\right\rangle , \ \ \
		\hbar\tilde{\omega}_k=\hbar\omega_0(k)+\frac{2\hbar e}{M_ec}\left\langle \phi\left|\vec{k}\vec{A}_R\right|\phi\right\rangle.
\end{eqnarray}

In what follows in this section we will assume $\hbar=1$, $\omega_0(k)=\omega_0=1$, $M_e=1$.
Equation \eqref{10} suggests that the bipolaron Hamiltonian differs from the polaron one in that in the latter the
quantity $V_k$ is replaced by $\bar{V}_k$ and $\bar{T}$, $\bar{U}$, $\bar{\Pi}$ are added to the polaron Hamiltonian.

With the use of Lee-Low-Pines canonical transformation~\cite{43}
\begin{equation*}
    S_2=\text{exp}\left\{\sum_kf_k(a^+_k-a_k)\right\},
\end{equation*}
where $f_k$ are variational parameters with the sense of the distance by which the field oscillators are displaced from their equilibrium positions:

\begin{equation*}
    S^{-1}_2a_kS_2=a_k+f_k,\ \ \ S^{-1}_2a^+_kS_2=a^+_k+f_k,
\end{equation*}
for Hamiltonian $\tilde{\tilde{H}}$:

\begin{equation}\label{14}
    \tilde{\tilde{H}}=S^{-1}_2\bar{H}S_2,
\end{equation}
$$ \tilde{\tilde{H}}=H_0+H_1,$$
we get:
\begin{multline}\label{15}
    H_0=2\sum_k\bar{V}_kf_k+\sum_kf_k^2\tilde{\omega}_k+\frac{1}{2}\left(\sum_k\vec{k}f^2_k\right)^2+
		\mathcal{H}_0+\bar{T}+\bar{U}+\bar{\Pi},\\
		 \mathcal{H}_0=\sum_k\omega_ka^+_ka_k+\frac{1}{2}\sum_{k,k'}\vec{k}\vec{k'}f_kf_{k'}
		(a_ka_{k'}+a^+_ka^+_{k'}+a^+_ka_{k'}+a^+_{k'}a_k),
\end{multline}
where:

\begin{equation}\label{16}
    \omega_k=\tilde{\omega}_k+\frac{k^2}{2}+\vec{k}\sum_{k'}\vec{k}'f^2_{k'}.
\end{equation}

Hamiltonian $H_1$ contains terms linear, threefold and fourfold in the creation and annihilation operators.
Its explicit form is given in~\cite{38,40}.

Then, as is shown in~\cite{38,40}, the use of Bogolyubov-Tyablikov canonical transformation~\cite{44}
for passing on from operators $a^+_k$, $a_k$ to new operators $\alpha ^+_k$, $\alpha _k$:
$$a_k=\sum_{k'}M_{1kk'}\alpha_{k'}+\sum_{k'}M^*_{2kk'}\alpha^+_{k'}$$

\begin{equation}\label{17}
     a^+_k=\sum_{k'}M^*_{1kk'}\alpha^+_{k'}+\sum_{k'}M_{2kk'}\alpha_{k'},
\end{equation}
(in which $\mathcal{H}_0$ is a diagonal operator) makes mathematical expectation of $H_1$ equal to zero
in the absence of magnetic field (see Appendix). The contribution of $H_1$ to the spectrum of transformed Hamiltonian when magnetic field is non-zero is discussed in Section~\ref{Energy}.

In the new operators $\alpha ^+_k$, $\alpha _k$ Hamiltonian \eqref{15} takes on the form
$$\tilde{\tilde{\tilde{H}}}=E_{bp}+\sum_k\nu_k\alpha^+_k\alpha_k,$$
\begin{equation}\label{18}
    E_{bp}=\Delta E_r+2\sum_k\bar{V}_kf_k+\sum_k\tilde{\omega}_kf^2_k+\bar{T}+\bar{U}+\bar{\Pi},
\end{equation}
where $\Delta E_r$ is the so-called recoil energy.
The general expression for $\Delta E_r=\Delta E   _r\left\{f_k\right\}$ was obtained in~\cite{40}.
Actually, calculation of the ground state energy $E_{bp}$ for the case of Froehlich Hamiltonian was performed in~\cite{38,40} by minimization of \eqref{18}
in $f_k$ and in $\Psi$ in the absence of a magnetic field.

Notice that in the theory of a polaron with broken symmetry a diagonalized electron-phonon Hamiltonian has the form of \eqref{18}~\cite{44-Miyake}. This Hamiltonian can be treated as a Hamiltonian of a polaron and a system of its associated renormalized real phonons or as a Hamiltonian whose quasiparticle excitations spectrum is determined by \eqref{18}~\cite{45-Levenson}. In the latter case excited states of a polaron are Fermi quasiparticles.

In the case of a bipolaron the situation is qualitatively different since a bipolaron is a boson quasiparticle whose spectrum is determined by \eqref{18}. Obviously, a gas of such quasiparticles can experience Bose-Einstein condensation (BEC). Treatment of \eqref{18} as a bipolaron and its associated renormalized phonons does not prevent their BEC since maintenance of the number of particles required in this case takes place automatically due to commutation of the total number of renormalized phonons with Hamiltonian \eqref{18}.

Renormalized frequencies $\nu _k$ involved in \eqref{18}, according to~\cite{38,40,45} are determined by the equation for $s$:

\begin{equation}\label{19}
    1=\frac{2}{3}\sum_k\frac{k^2f^2_k\omega_k}{s-\omega^2_k},
\end{equation}
solutions of which yield the spectrum of $s=\left\{\nu^2_k\right\}$ values. This equation has a general form and can be applied for arbitrary dependence of $V_k$ and $\omega_k$ on $k$.

\section{Energy Spectrum of a TI-Bipolaron}\label{Energy}

Hamiltonian \eqref{18} is conveniently presented in the form:

\begin{equation}\label{20}
    \tilde{\tilde{\tilde{H}}}=\sum_{n=0,1,2,...}E_{n}\alpha^+_{n}\alpha_{n},
\end{equation}
\begin{equation}\label{21}
    E_{n}=\left\{\begin{array}{rl}
		E_{bp},\ n=0;\\
		\nu_{n}=E_{bp}+\omega_{k_n},\ n\neq0 ;
		\end{array}\right. \
\end{equation}
where in the case of a three-dimensional ionic crystal $\vec{k}_n$ is a vector with the components:
\begin{equation*}\label{22}
    k_{n_i}=\pm\frac{2\pi(n_i-1)}{N_{a_i}},\ \ n_i=1,2,...,\frac{N_{a_i}}{2}+1,\ i=x,y,z,
\end{equation*}

$N_{ai}$ is the number of atoms along the $i$-th crystallographic axis.

Let us prove the validity of the expression for the spectrum \eqref{20}, \eqref{21}. Since operators $\alpha ^+_n$,
$\alpha _n$ obey Bose commutation relations:

\begin{equation*}\label{23}
    \left[\alpha_{n},\alpha^+_{n'}\right]=\alpha_{n}\alpha^+_{n'}-\alpha^+_{n'}\alpha_{n}=
		\delta_{n,n'},
\end{equation*}
they can be considered to be operators of creation and annihilation of TI-bipolarons.
The energy spectrum of TI-bipolarons, according to \eqref{19}, is determined by the equation:

\begin{equation}\label{24}
    F(s)=1,
\end{equation}
where:

\begin{equation*}\label{25}
    F(s)=\frac{2}3{\sum_{n}}\frac{k^2_{n}f^2_{k_n}\omega^2_{k_n}}{s-\omega^2_{k_n}}.
\end{equation*}

\begin{figure}
\centering
\includegraphics[width=\linewidth]{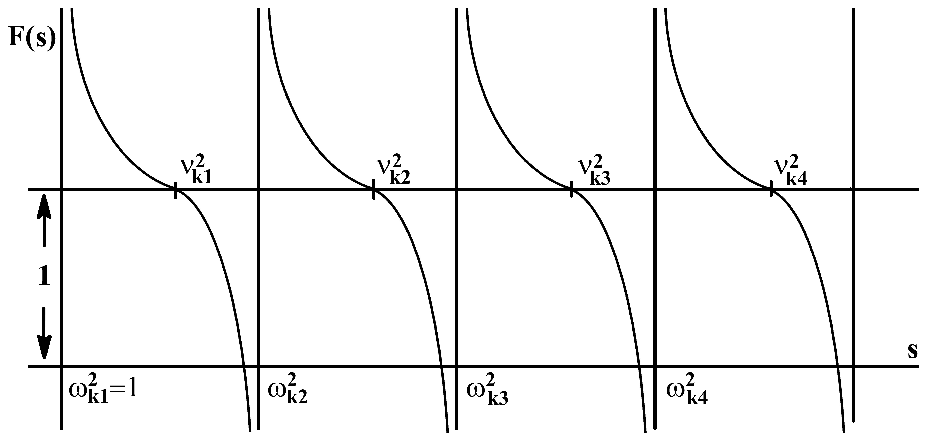}
\caption{Graphical solution of Equation \eqref{24}.}\label{fig1}
\end{figure}

It is convenient to solve Equation \eqref{24} graphically (Figure~\ref{fig1}).

Figure~\ref{fig1}  suggests that the frequencies $\nu_{k_n}$   lie between the frequencies $\omega_{k_n}$ and $\omega_{k_{n+1}}$.
Hence, the spectrum $\nu_{k_n}$ as well as the spectrum $\omega_{k_n}$ are quasicontinuous:
$\nu_{k_n}-\omega_{k_n}=O(N^{-1})$ which just proves the validity of \eqref{20} and \eqref{21}.

It follows that the spectrum of a TI-bipolaron has a gap between the ground state $E_{bp}$
 and the quasicontinuous spectrum, equal to $\omega_0$.

In the absence of an external magnetic field, functions $f_k$ involved in expression for $\omega _k$ \eqref{16} are independent of the direction of the wave vector $\vec{k}$. When an external magnetic field is applied, $f_k$ cannot be considered to be an isotropic quantity, accordingly, we cannot put the last term in equation \eqref{16} for $\omega _k$ equal to zero. Besides, the angular dependence involved in the spectrum $\omega _k$ in the magnetic field is also contained in the term $\tilde{\omega} _k$
 involved in the quantity $\omega _k$. Since in the isotropic system discussed there is only one preferred direction determined by vector $\vec{B}$, for $\omega _k$ from \eqref{16} we will get:

\begin{equation}\label{26}
    \omega _{k_n}=\omega_0+\frac{\hbar k^2_n}{2M_e}+\frac{\eta}{M_e}\left(\vec{B}\vec{k}_n\right),
\end{equation}
where $\eta$ is a scalar quantity. Notice that the contribution of $H_1$ to spectrum \eqref{26}
will lead to the dependence of $\eta$ from $|\vec{k}|$ and $(\vec{k}\vec{B})$.
For weak magnetic field in longitudinal limit (when Froehlich Hamiltonian is valid)
we will neglect such dependence and consider $\eta$ as a constant value.

For a magnetic field directed along the axis $z$, expression \eqref{26} can be written as:

\begin{equation}\label{27}
    \omega _{k_n}=\omega_0+\frac{\hbar^2}{2M_e}\left(k_{zn}+k^0_z\right)^2+
		\frac{\hbar^2}{2M_{e}}\left(k^2_{xn}+k^2_{yn}\right)^2-
		\frac{\eta^2B^2}{2\hbar^2M_e}.
\end{equation}

Note that formula \eqref{27} can be generalized to the anisotropic case when in the directions $k_x$ and $k_y$:
$M_{ex}=M_{ey}=M_{||}$, and in the direction $k_z$: $M_{ez}=M_{\bot}$(Section~\ref{Current}).
Formula \eqref{27} in this case takes on the form:

\begin{align}\tag{\ref{27}$\,'$}\label{27a} 
\omega _{k_n}=\omega_0+\frac{\hbar^2}{2M_{\bot}}\left(k_{zn}+k^0_z\right)^2+
		\frac{\hbar^2}{2M_{||}}\left(k^2_{xn}+k^2_{yn}\right)-
		\frac{\eta^2B^2}{2\hbar^2M_{\bot}},
\end{align}		

if the magnetic field is directed along the axis $z$ and:
\begin{align}\tag{\ref{27}$\,''$}\label{27b}
\omega _{k_n}=\omega_0+\frac{\hbar^2}{2M_{\bot}}k_{zn}^2+
		\frac{\hbar^2}{2M_{||}}\left(k_{xn}+k^0_{xn}\right)^2 +\frac{\hbar^2}{2M_{||}}k^2_{yn}-
		\frac{\eta^2B^2}{2\hbar^2M_{||}},
\end{align}		
if the magnetic field is directed along the axis $x$.	

Below we will consider the case of low concentration of TI-bipolarons in a crystal.
Then they can adequately be considered to be an ideal Bose gas, whose properties are determined by Hamiltonian~\eqref{20}.	

\section{Statistical Thermodynamics of Low-Density TI Bipolarons without Magnetic Field}\label{Stat}
Let us consider an ideal Bose-gas of TI-bipolarons which represents a system of $N$
particles occurring in some volume $V$. Let us write $N_0$ for the number of particles
in the lower one-particle state and $N'$ for the number of particles in higher states. Then:

\begin{equation}\label{28}
    N=\sum_{n=0,1,2,...}\bar{m}_{n}=\sum_{n}\frac{1}{e^{(E_{n}-\mu)/T}-1},
\end{equation}
or:
\begin{equation}\label{29}
    N=N_0+N',\ \ N_0=\frac{1}{e^{(E_{bp}-\mu)/T}-1},\ \
		N'=\sum_{n\neq0}\frac{1}{e^{(E_{n}-\mu)/T}-1}.
\end{equation}

\begin{figure}
\begin{center}
\includegraphics[width=\linewidth]{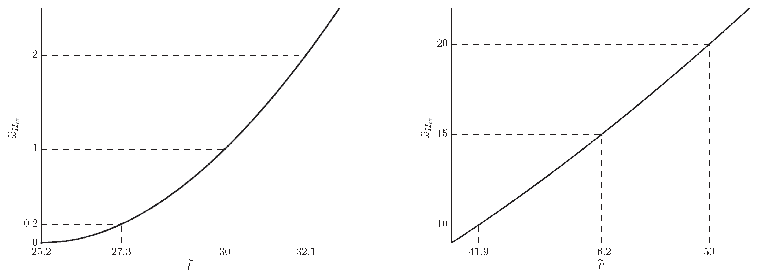}
\caption{Solutions
of Equation \eqref{30} with $C_{bp}=331.35$ and $\tilde{\omega}_i=\left\{0.2; 1; 2; 10; 15; 20\right\}$,
which correspond to  $\tilde{T}_{c_i}$: $\tilde{T}_{c_1}=27.3$; $\tilde{T}_{c_2}=30$;
$\tilde{T}_{c_3}=32.1$; $\tilde{T}_{c_4}=41.9$; $\tilde{T}_{c_5}=46.2$; $\tilde{T}_{c_6}=50$.}\label{fig2}
\end{center}
\end{figure}

In expression $N'$ \eqref{29} we will perform integration over quasicontinuous spectrum (instead of summation)
\eqref{20}, \eqref{21} and \eqref{27} and assume $\mu=E_{bp}$. As a result, from \eqref{28} and \eqref{29} we get an equation for determining the critical temperature of Bose-condensation $T_c$:

\begin{equation}\label{30}
    C_{bp}=f_{\tilde{\omega}_H}\left(\tilde{T}_c\right),
\end{equation}
$$f_{\tilde{\omega}_H}\left(\tilde{T}_c\right)=\tilde{T}^{3/2}_cF_{3/2}\left(\tilde{\omega}_H/\tilde{T}_c\right),\ \
F_{3/2}(\alpha)=\frac{2}{\sqrt{\pi}}\int^{\infty}_0\frac{x^{1/2}dx}{e^{x+\alpha}-1},$$
$$C_{bp}=\left(\frac{n^{2/3}2\pi\hbar^2}{M_e\omega^*}\right)^{3/2},\ \
\tilde{\omega}_H=\frac{\omega_0-\eta^2H^2/2M_e}{\omega^*},\ \ \tilde{T}_c=\frac{T_c}{\omega^*},$$
where $n=N/V$. In this section we will deal with the case when the magnetic field is lacking: $H=0$.
Figure~\ref{fig2} shows a graphical solution of Equation \eqref{30} for the values of parameters $M_e=2m^*=2m_0$,
where  $m_0$ is the mass of a free electron in vacuum, $\omega ^*=5$ meV ($\approx $58 K), $n=10^{21}$ cm$^{-3}$
and the values: $\tilde{\omega}_1=0.2$; $\tilde{\omega}_2=1$; $\tilde{\omega}_3=2$; $\tilde{\omega}_4=10$;
$\tilde{\omega}_5=15$; $\tilde{\omega}_6=20$; $\tilde{\omega}_H=\tilde{\omega}=\omega _0/\omega ^*$.

It is seen from Figure~\ref{fig2} that the critical temperature grows with increasing phonon frequency $\omega _0$.
It is also evident from Figure~\ref{fig2} that an increase in the concentration
of TI-bipolarons $n$ will lead to an increase in the critical temperature, while a gain in the electron mass
$m^*$ to its decrease.  For $\tilde{\omega}=0$ the results go over into the limit of ideal Bose-gas (IBG).
In particular, \eqref{30} for $\tilde{\omega}=0$, yields the expression for the critical temperature of IBG:

\begin{equation}\label{31}
    T_c=3.31\hbar^2n^{2/3}/M_e.
\end{equation}

It should be stressed, however, that \eqref{31} involves $M_e=2m^*$, rather than the bipolaron mass.
This resolves the problem of the low temperature of condensation which arises both in the small radius polaron
theory and in the large radius polaron  theory in which expression \eqref{31} involves the bipolaron mass~\cite{46,47,48,49,50,51,52,52a}.
Another important result is that the critical temperature $T_c$ for the parameter values
considerably exceeds the gap energy $\omega _0$. Such values as energy, free energy, heat capacity and the transition heat of TI-bipolaron gas were calculated for the case $H=0$ in \cite{Lakhno-ACMP,lakhno-physC}.

\section{Current States of a TI-Bipolaron Gas}\label{Current}		

As is known, the absence of a magnetic field inside a superconductor is caused
by the existence of surface currents compensating this field.
Thus, from condition \eqref{8} it follows that:

\begin{equation}\label{39}
		\vec{\mathcal{P}}_R=-\frac{2e}{c}\vec{A}_R,
\end{equation}
i.e., in a superconductor there is a persistent current $\vec{j}$:

\begin{equation}\label{40}
		\vec{j}=2en_0\vec{\mathcal{P}}_R/M^*_e=-\frac{4e^2n_0}{M^*_ec}\vec{A}_R,
\end{equation}
(where $M^*_e$ is bipolaron effective mass), providing Meissner effect, where $n_0$ is a concentration of superconducting charge carriers:
$n_0=N_0/V$. Comparing \eqref{40} with the well-known phenomenological expression for the surface current $\vec{j}_S$~\cite{53}:

\begin{equation}\label{41}
		\vec{j}_S=-\frac{c}{4\pi\lambda^2}\vec{A},
\end{equation}
and putting $\vec{A}=\vec{A}_R$, with the use of \eqref{40}, \eqref{41} and equality $\vec{j}=\vec{j}_S$
 we will get a well-known expression for London penetration depth $\lambda$:

\begin{align}\label{42}
		\lambda=\left(\frac{M^*_ec^2}{16\pi e^2n_0}\right)^{1/2},
\end{align}

The equality of 'microscopic' expression for current \eqref{40} to its 'macroscopic' value cannot be exact.
Accordingly, the equality $\vec{A}=\vec{A}_R$ is also approximate since $\vec{A}_R$
represents a vector-potential at the point where the center of mass of two electrons occurs, while in London theory
$\vec{A}$ is a vector-potential at the point where a particle resides. For this reason, these two quantities should better be considered proportional. In this case the expression for the penetration depth has the form:

\begin{align}\label{42a}
\lambda=\text{const}\left(\frac{M^*_ec^2}{16\pi e^2n_0}\right)^{1/2}, \tag {$30'$}
\end{align}
where the constant multiplier in \eqref{42a} (of the order of unity) should be determined from a comparison with the experiment.

Expression \eqref{39} was obtained in the case of isotropic effective mass of charge carriers.
However, actually, it has a more general character and does not change when anisotropy of
effective masses is taken into account. Thus, for example in layered HTSC materials kinetic energy
of charge carriers in Hamiltonian \eqref{1} should be replaced by the expression:

\begin{eqnarray*}\label{43}
     T_a=\frac{1}{2m^*_{||}} {\left(\hat{P}_{1||}-\frac{e}{c}\vec{A}_1\right)^2}+
		\frac{1}{2m^*_{||}} {\left(\hat{P}_{2||}-\frac{e}{c}\vec{A}_2\right)^2}+\\ \nonumber
		\frac{1}{2m^*_{\bot}} {\left(\hat{P}_{1\bot}-\frac{e}{c}\vec{A}_{1z}\right)^2}+
		\frac{1}{2m^*_{\bot}} {\left(\hat{P}_{2\bot}-\frac{e}{c}\vec{A}_{1z}\right)^2},
\end{eqnarray*}
where $\hat{P}_{1,2||}$, $\vec{A}_{1,2||}$ are operators of the momentum and vector-potential in the planes of layers
(ab planes); $\hat{P}_{1,2\bot}$, $\vec{A}_{1,2\bot}$ are relevant quantities in the direction
perpendicular to the planes (along c-axis); $m^*_{||}$, $m^*_{\bot}$
are effective masses in the planes and in the perpendicular direction.

As a result of transformation:

\begin{eqnarray}\label{44}
		\tilde{x}=x,\ \ \ \ \ \tilde{y}=y,\ \ \ \ \ \tilde{z}=\gamma z \\ \nonumber
		\tilde{A}_{\tilde{x}}=A_x,\ \ \tilde{A}_{\tilde{y}}=A_y,\ \ \tilde{A}_{\tilde{z}}=\gamma ^{-1}A_z,\\ \nonumber
		\hat{\tilde{\mathcal{P}}}_{\tilde{x}}=\hat{\mathcal{P}}_x,\ \
		\hat{\tilde{\mathcal{P}}}_{\tilde{y}}=\mathcal{P}_y,\ \
		\tilde{\mathcal{P}}_{\tilde{z}}=\gamma^{-1}\mathcal{P}_z,
\end{eqnarray}
where $\gamma^2=m^*_{\bot}/m^*_{||}$, $\gamma$ is the anisotropy parameter
kinetic energy $\tilde{T}_a$ appears to be isotropic. It follows that:
$\hat{\tilde{\mathcal{P}}}_R+(2e/c)\vec{\tilde{A}}_{\tilde{R}}=0$.
Then \eqref{44} suggests that relation \eqref{39} appears to be valid in the anisotropic case too. It follows that:

$$\vec{\mathcal{P}}_{R||}=-\frac{2e}{c}\vec{A}_{R||},\ \vec{\mathcal{P}}_{R\bot}=-\frac{2e}{c}\vec{A}_{R\bot}, $$

\begin{equation*}\label{45}
		\vec{j}_{||}=2en_0\mathcal{P}_{R||}/M^*_{e||},\ \ \
		\vec{j}_{\bot}=2en_0\vec{\mathcal{P}}_{R\bot}/M^*_{e\bot}.
\end{equation*}

The magnetic field directed perpendicular to the plane of layers will induce currents running in the plane of layers.
Having penetrated into a sample, such a field will decrease along the plane of layers.
Let us write $\lambda_{||}$  for the London penetration depth of the magnetic field perpendicular to the plane of layers
($H_{\bot}$) and $\lambda_{\bot}$ for that of the magnetic field parallel to the plane of layers ($H_{||}$).

This suggests expressions for London depths of the magnetic field penetration into a sample:

\begin{equation}\label{46}
		\lambda _{\bot}=\left(\frac{M^*_{e\bot}c^2}{16\pi e^2n_0}\right)^{1/2},
		\lambda _{||}=\left(\frac{M^*_{e||}c^2}{16\pi e^2n_0}\right)^{1/2}.
\end{equation}

For $\lambda_{||}$ and $\lambda_{\bot}$, designations $\lambda_{ab}$ and $\lambda_{c}$ are also used.
From \eqref{46} it follows that:

\begin{equation}\label{47}
		\frac{\lambda _{\bot}}{\lambda _{||}}
		=\left(\frac{M^*_{e\bot}}{M^*_{e||}}\right)^{1/2}=\gamma^* .
\end{equation}

From \eqref{46} it also follows that the London penetration depth depends on temperature:

\begin{equation}\label{48}
		\lambda ^2(0)/\lambda^2(T)=n_0(T)/n_0(0).
\end{equation}

In particular, for $\omega=0$, with the use of \eqref{31} we get: $\lambda(T)=\lambda(0)\left(1-(T/T_C)^{3/2}\right)^{-1/2}$.
Comparison of the dependence obtained with those derived within other approaches is given in Section~\ref{Comparison}.

It is generally accepted that the Bose system became superconducting due to the inter-particle interaction. The existence of a gap in TI-bipolaron spectrum can drive their condensation and the Landau superfluidity condition:

\begin{equation}\label{48-1}
		v<\hbar\omega_0/\mathcal{P}\ .
\end{equation}
(where $\mathcal{P}$ is the momentum of bipolaron condensate)
can be fulfilled even for noninteracting particles. From condition \eqref{48-1} it follows the expression for maximum value of current density $j_{\max}=env_{\max}$:

\begin{equation*}\label{48-2}
        j_{\max}=en_0\sqrt{\frac{\hbar\omega_0}{M^*_e}}
\end{equation*}

It should be noted that all the aforesaid refers to local electrodynamics.
Accordingly, expressions obtained for $\lambda$ are valid only on condition that $\lambda>>\xi$ ,
where $\xi$ is a correlation length determining the characteristic size of the pair,
i.e., the characteristic scale of changes of the wave function $\psi(r)$ in \eqref{9}.
This condition is usually fulfilled in HTSC materials.
In ordinary superconductors the inverse inequality is valid.
Nonlocal generalization of superconductor electrodynamics was made by Pippard~\cite{54}.
Within this approach relation between $\vec{j}_S$ and $\vec{A}$ in expression \eqref{41} can be written as:

\begin{equation*}\label{49}
		\vec{j}_S=\int\hat{Q}(z-r')\vec{A}(r')d\vec{r}',
\end{equation*}
where $Q$ is a certain operator whose radius of action is usually believed to be equal to $\xi$.
In the limit of $\xi>>\lambda$ this leads to an increase in the absolute value of the length of the magnetic field penetration into a superconductor which becomes equal to $(\lambda^2\xi)^{1/3}$~\cite{53}.

\section{Thermodynamic Properties of a TI-Bipolaron Gas in a Magnetic Field}\label{Thermodyn}

To start with, let us notice that expression for $\tilde{\omega}_H$ \eqref{30} suggests that for $\omega_0=0$
Bose-condensation appears to be impossible if $H\neq0$.
For an ordinary ideal charged Bose-gas, this conclusion was first made in~\cite{55}.
In view of the fact that in the spectrum of TI-bipolarons
there is a gap between the ground state of a TI-bipolaron gas and
the excited one (Section~\ref{Energy}), this conclusion becomes invalid for $\omega_0\neq0$.

Expression $\tilde{\omega}_H$ \eqref{30} suggests that there is a maximum value of the magnetic field $H_{max}$ equal to:

\begin{equation}\label{50}
		H^2_{max}=\frac{2\omega_0\hbar^2M_e}{\eta ^2}.
\end{equation}

As follows from \eqref{16}, the value $\eta$ consists from two parts: $\eta=\eta'+\eta''$. The value $\eta'$ is determined by the integral entering into the expression for  $\tilde{\omega}_k$ \eqref{11}. For this reason $\eta'$ depends on the form of sample surface. The value $\eta''$ is determined by the sum entering into the expression for $\omega_k$ \eqref{16} and weakly depends on surface form. This leads to the conclusion that the value $\eta$ can be changed by changing sample surface and thus changing $H_{max}$. For $H>H_{max}$ a homogeneous superconducting state is impossible.
With the use of \eqref{50}, $\tilde{\omega}_H$ \eqref{30} will be written as:

\begin{equation}\label{51}
		\tilde{\omega}_H=\tilde{\omega}\left(1-H^2/H^2_{max}\right).
\end{equation}

For a given temperature $T$, let us write $H_{cr}(T)$ for the value of the magnetic field
at which the superconductivity disappears.
According to \eqref{51}, this value of the field corresponds to $\tilde{\omega}_{H_{cr}}$:
\begin{equation}\label{52}
		\tilde{\omega}_{Hcr}(T)=\tilde{\omega}\left(1-H^2_{cr}(T)/H^2_{max}\right).
\end{equation}

The temperature dependence of the quantity $\tilde{\omega}_{H_{cr}}(T)$ can be found from Equation \eqref{30}:
\begin{equation*}\label{53}
		 C_{bp}=\tilde{T}^{3/2}F_{3/2}\left(\tilde{\omega}_{Hcr}(\tilde{T})/\tilde{T}\right).
\end{equation*}

It has the form given in Figure~\ref{fig2} if we replace $\tilde{\omega}$ by $\tilde{\omega}_{H_{cr}}$ and $\tilde{T}_{c}$ by $\tilde{T}$.

Using \eqref{52} and the temperature dependence given in Figure~\ref{fig2} we can find the temperature dependence of $H_{cr}(\tilde{T})$:
\begin{equation}\label{54}
		\frac{H^2_{cr}(\tilde{T})}{H^2_{max}}=1-\omega_{Hcr}(\tilde{T})/\tilde{\omega}.
\end{equation}

\begin{figure}
\begin{center}
\includegraphics[width=0.8\linewidth]{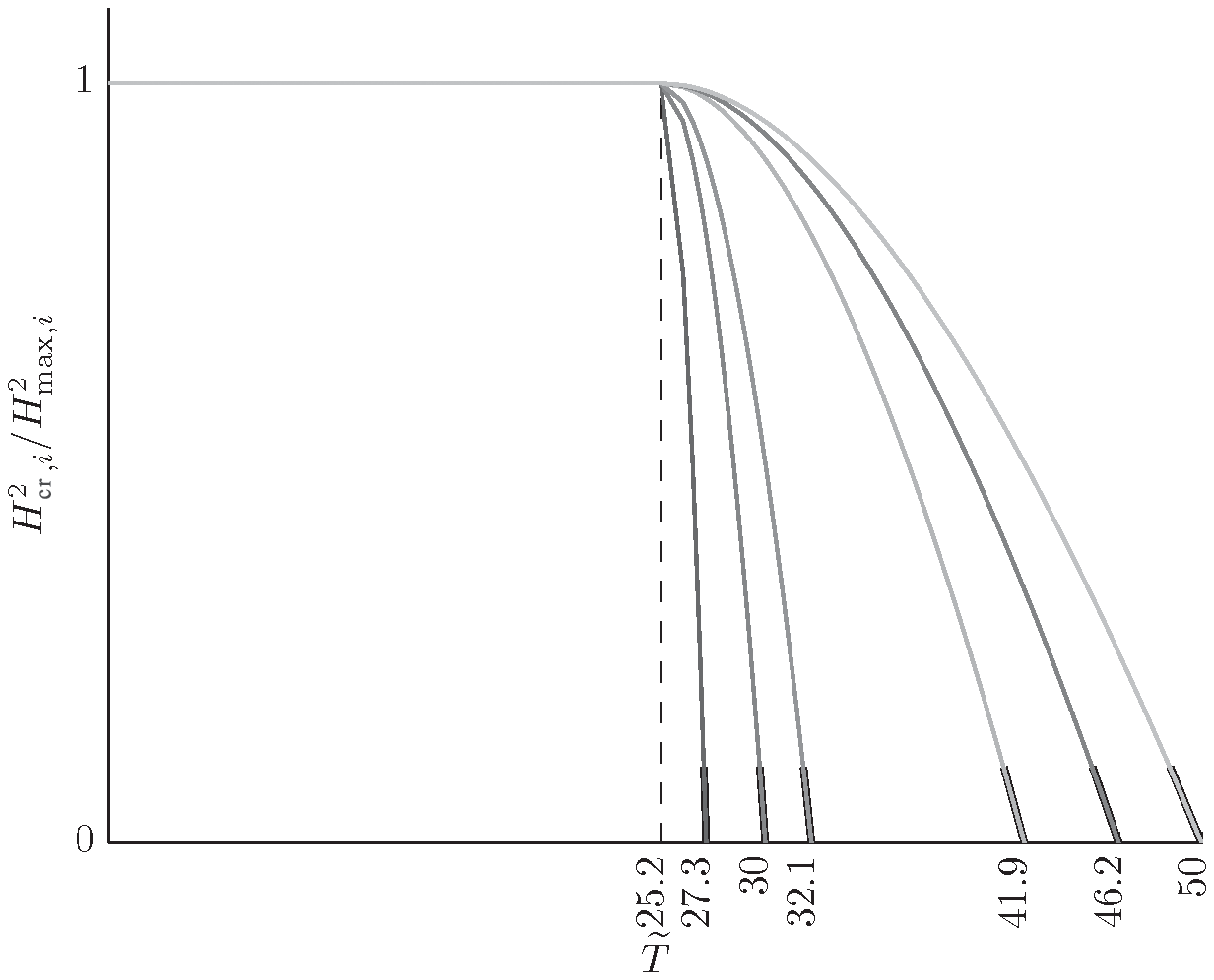}
\caption{Temperature dependencies $H^2_{cr,i}/H^2_{max,I}$ on the intervals $[0;T_{c,i}]$
for the values of parameters $\tilde{\omega}_i$, given in Figure~\ref{fig2}.}\label{fig6}
\end{center}
\end{figure}

For $\tilde{T}\leq\tilde{T}_{ci}$, these dependencies are given in Figure~\ref{fig6}.

As is seen from Figure~\ref{fig6}, $H_{cr}(\tilde{T})$ reaches its maximum at a finite temperature of
$\tilde{T}_c(\tilde{\omega}=0)\leq\tilde{T}_c(\omega_{0i})$.
Figure~\ref{fig6} suggests that at temperature below $\tilde{T}_c(\tilde{\omega}=0)=25.2$ a further decrease of the temperature
no longer changes the value of the critical field $H_{cr}(\tilde{T})$ irrespective of the gap value $\tilde{\omega}$.

\begin{figure}
\begin{center}
\includegraphics[width=0.8\linewidth]{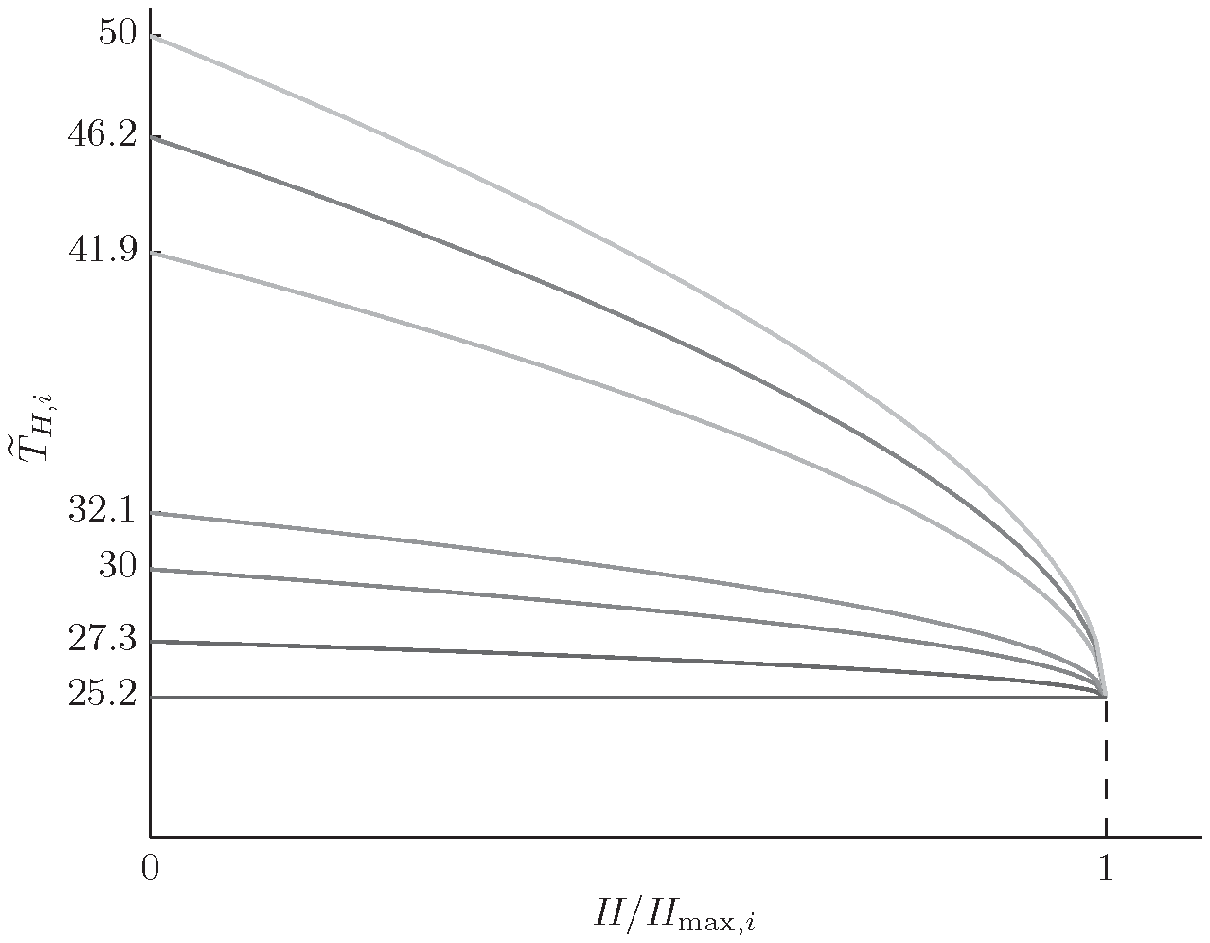}
\caption{Dependencies of the critical transition temperature $\tilde{T}_{H,i}$ on the magnetic field $H$
for the values of parameters $\tilde{\omega}_i$ given in Figure~\ref{fig2}.}\label{fig7}
\end{center}
\end{figure}

Let us also introduce the notion of a transition temperature $T_c(H)$ in the magnetic field $H$.
Figure~\ref{fig7} illustrates the dependencies $T_c(H)$ resulting from Figure~\ref{fig6} and determined by the relations:
\begin{equation*}\label{55}
		C_{bp}=\tilde{T}^{3/2}F_{3/2}\left(\tilde{\omega}_{H,i}/\tilde{T}_{C,i}(H)\right),\ \ \
		\tilde{\omega}_{H,i}=\tilde{\omega}_{H=0,i}\left[1-H^2/H^2_{max,i}\right].
\end{equation*}

Figure~\ref{fig7} suggests that the critical transition temperature $\tilde{T}_{c}(H)$ changes stepwise as
the magnetic field reaches the value $H_{max,i}$.

To solve the problem of the type of the phase transition in a magnetic field let us proceed from the well-known expression which relates free energies in superconducting and normal states:

\begin{equation}\label{56}
		F_S+\frac{H^2}{8\pi}=F_N,
\end{equation}
where $F_S$ and $F_N$ are free energies of the unit volume of superconducting and normal states, respectively:

\begin{equation*}\label{57}
		F_ S=\frac{N}{V}E_{bp}(H=0)-\frac{2}{3}\Delta E(\omega_{H=0})\frac{N}{V},
\end{equation*}

\begin{equation*}\label{58}
		F_N=\frac{N}{V}E_{bp}(H)-\frac{2}{3}\Delta E(\omega_{H})\frac{N}{V},
\end{equation*}
where $\Delta E=E-E_{bp}$, $E=\omega^*\tilde{E}$, where $\tilde{E}$ is determined by formula (29) in \cite{Lakhno-ACMP}. 
Differentiating \eqref{56} with respect to temperature and taking into account that $S=-\partial F/\partial T$, we express the heat of transition $q$~as:

\begin{equation}\label{59}
		q=T(S_N-S_S)=-T\partial(F_N-F_S)/\partial T=
		-T\frac{H_{cr}}{4\pi}\frac{\partial H_{cr}}{\partial T},
\end{equation}

Accordingly the difference of entropies $S_S-S_N$ will be written as:

\begin{equation}\label{60}
		S_S-S_N=\frac{H_{cr}}{4\pi}\left(\frac{\partial H_{cr}}{\partial T}\right)=\frac{H^2_{max}}{8\pi\omega^*}
		(\tilde{S}_S-\tilde{S}_N).
\end{equation}

\begin{figure}
\centering
\includegraphics[width=0.75\linewidth]{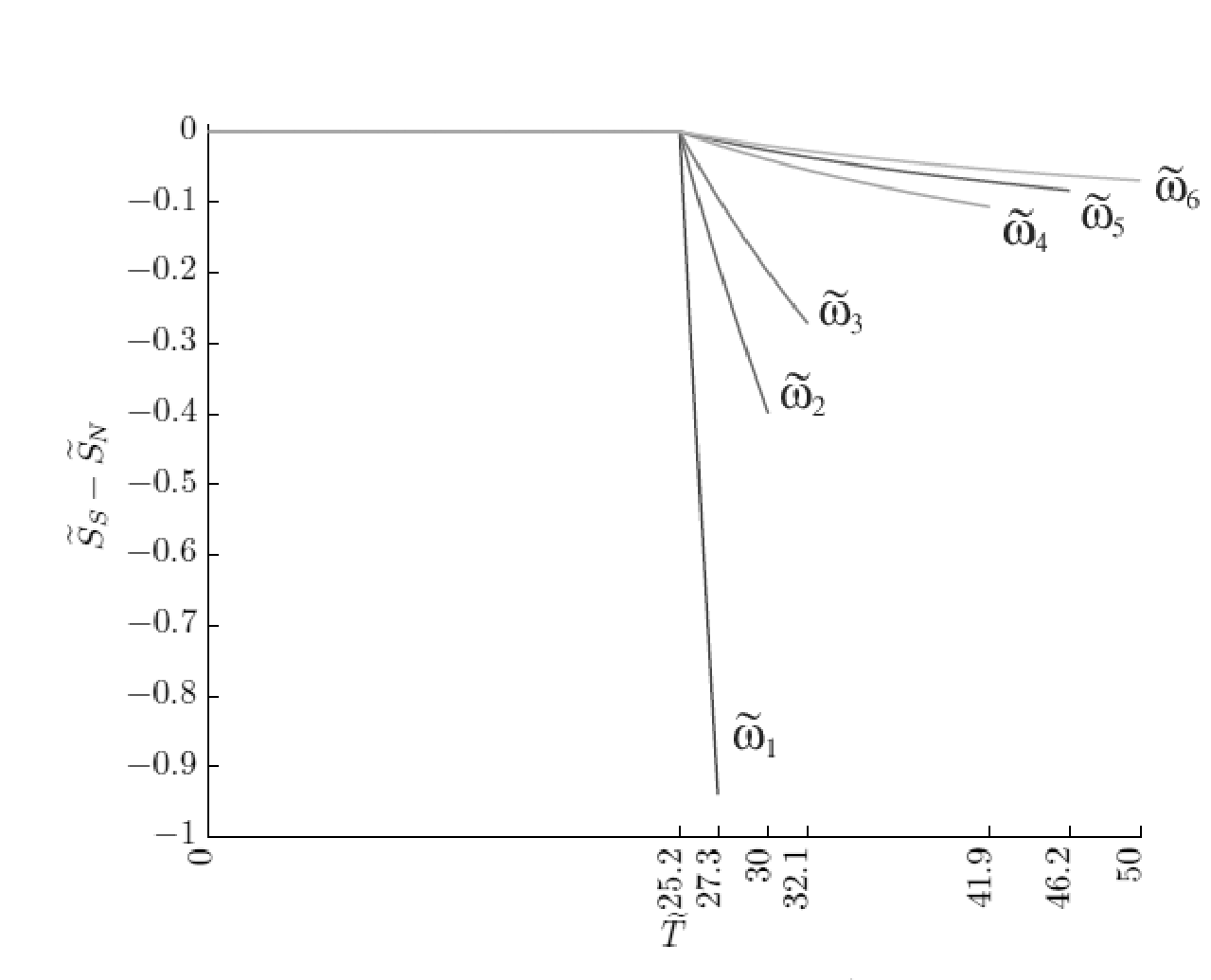}
\caption{Temperature dependencies for the difference of entropies of superconducting and
normal states for the values of parameters  $\tilde{\omega}_i$ given in Figures~\ref{fig6} and  \ref{fig7}.}\label{fig8}
\end{figure}

Figure~\ref{fig8} shows the temperature dependence of the difference of entropies \eqref{60} for various values
of critical temperatures $(\tilde{\omega}_i)$ given in Figure~\ref{fig2}. These dependencies may seem strange in, at least, two respects:
\begin{enumerate}
\item	In BCS and Ginzburg-Landau theory at the most critical point $T_c$ the difference of entropies becomes zero in accordance with Rutgers formula.
In Figure~\ref{fig8} entropy is a monotonous function $\tilde{T}$ which does not vanish for $T=T_c$.
\item	Second, in absolute terms, the difference $|\tilde{S}_S-\tilde{S}_N|$, when approaching the limit point
$\tilde{T}_c=25.2$, which corresponds to the value $\tilde{\omega}=0$, as it can be seemed
should decrease rather than increase vanishing at $\tilde{\omega}=0$.
\end{enumerate}

As for the second point, this is really the case for $|S_S-S_N|$,
since the value of the maximum field $H_{max}$ and, accordingly, the multiplier $H^2_{max}/8\pi$
relating the quantities $S_S-S_N$ and $\tilde{S}_S-\tilde{S}_N$ becomes zero for $\tilde{\omega}=0$.

As for the first point, as will be shown below, Rutgers formula appears inapplicable for Bose-condensate of TI-bipolarons.

Table~\ref{table2} lists the values of the quantity $\tilde{S}_S-\tilde{S}_N$ for critical
temperatures corresponding to different values of $\tilde{\omega}_{H_{cr,i}}$.

The results obtained suggest some fundamental conclusions:
\begin{enumerate}
\item	The curve of the dependence $H_{cr}(T)$ (Figure~\ref{fig6}) for $T=0$ has a zero derivative,
accordingly $dH_{cr}(T)/dT=0$ for $T=0$. This result is consistent with Nernst theorem which implies
that entropy determined by \eqref{59} is equal to zero for $T=0$.
\item	According to Figure~\ref{fig6}, $H_{cr}(T)$ is a curve monotonously drooping with increasing
$T$ for $T>T_c(\tilde{\omega}=0)$, and a constant value for $T\leq T_c(\tilde{\omega}=0)$.
Hence $\partial H_{cr}(T)/\partial T<0$ for \linebreak $T>T_c(\tilde{\omega}=0)$.
Therefore on the temperature interval $[T_c(\tilde{\omega}=0),T_c(\tilde{\omega})]$ $S_S<S_N$
and on the interval $[0,T_c(\tilde{\omega}=0)]$ $S_S=S_N$.
\end{enumerate}

This suggests some important conclusions:
\begin{enumerate}
\item	Transition on the interval $[0,T_c(\tilde{\omega}=0)]$ occurs without absorption or release
of latent heat since in this case $S_S=S_N$.
Experimentally it will be seen as a phase transition of the second kind.
Actually, in the region $[0,T_c(\tilde{\omega}=0)]$, a phase transition into a superconducting state is a phase transition
of infinite kind, since in this region, according to \eqref{56} and Figure~\ref{fig6},
any-order derivatives of the difference of free energies $F_S-F_N$, become zero.
\item	Passing in a magnetic field from a superconducting state to a normal one on the interval \linebreak $[T_c(\tilde{\omega}=0),T_c(\tilde{\omega})]$, which corresponds to $S_S<S_N$,
occurs with absorption of latent heat. On the contrary, passing from a normal state to
a superconducting one takes place with release of latent heat.
The phase transition on the interval $[0,T_c(\tilde{\omega}=0)]$ is not attended by absorption
or release of the latent heat being the phase transition of infinite kind.
\end{enumerate}

With regard to the fact that the specific heat capacity of a substance is determined
by the formula $C=T(\partial S/\partial T)$, the difference of specific heat capacities of superconducting
and normal states, according to \eqref{60} will be written as:

\begin{equation}\label{61}
		C_S-C_N=\frac{T}{4\pi}\left[\left(\frac{\partial H_{cr}}{\partial T}\right)^2+
		H_{cr}\frac{\partial^2H_{cr}}{\partial T^2}\right].
\end{equation}

This relation is usually used to get the well-known Rutgers formula.
To do this one assumes the critical field in \eqref{61} to be $H_{cr}(T_c)=0$ for  $T=T_c$
and leaves in the brackets on the right-hand side of \eqref{61} only the first term:

\begin{equation*}\label{62}
		\left(C_S-C_N\right)_R=\frac{T_c}{4\pi}\left(\frac{\partial H_{cr}}{\partial T}\right)^2_{T_c}.
\end{equation*}

It is easily seen, however, that at the point $T=T_c$ the quantity $\omega_{H_{cr}}$ determined by Figure~\ref{fig2},
for all the values of the temperature, has a finite derivative with respect to $T$ and,
therefore, according to \eqref{54}, an infinite derivative $\partial H_{cr}/\partial T$ for $T=T_c$.
Hence, the second term in the brackets \eqref{61} reduces to $-\infty$,
leaving this bracket a finite value. As a result, a proper expression for the difference of heat
capacities of the considered model of Bose-gas should be determined by the formula:

\begin{equation}\label{63}
		C_S-C_N=\frac{T}{8\pi}\frac{\partial ^2}{\partial T^2}H^2_{cr}(T)=\frac{H^2_{max}}{8\pi\omega^*}
		(\tilde{C}_S-\tilde{C}_N),
\end{equation}
$$\tilde{C}_S-\tilde{C}_N=\tilde{T}\frac{\partial ^2}{\partial \tilde{T}^2}\left(H^2_{cr}(\tilde{T})/H^2_{max}\right).$$

Table~\ref{table2} lists the values of quantity $\tilde{C}_S-\tilde{C}_N$ for the values of critical temperatures
corresponding to various values of $\tilde{\omega}_{H_{cr,i}}$. Notice that according to results obtained the capacity jump \eqref{63} has its maximum value at zero magnetic field and decreases as the magnetic field increases being equal zero at  $H=H_{cr}$ in full accordance with the experimental data \cite{Lakhno-ACMP}. Comparison of the jumps in the heat capacity
presented in \cite{Lakhno-ACMP} with expression \eqref{63} enables us co calculate the value of $H_{max}$.
The values of $H_{max}$ obtained by this means for various values of $\tilde{\omega}_i$ are given in Table~\ref{table2}.
These values unambiguously determine the values of constants $\eta$ in formulae \eqref{27a}, \eqref{27b}.

\begin{table}
	\centering
	\begin{tabular}{cccccc}
	\toprule
	\boldmath{$i$} & \boldmath{$\tilde{\omega}_{Hcr,i}$} & \boldmath{$\tilde{T}_{C_i}$} & \boldmath{$\tilde{S}_S-\tilde{S}_N$} & \boldmath{$\tilde{C}_S-\tilde{C}_V$} & \boldmath{$H_{max}\times 10^{-3}$\textbf{, Oe}} \\
	\midrule
		0  & 0 & 25.2 & 0 & 0 & 0 \\
		1  & 0.2 & 27.3 & $-$0.94 & $-$11.54 & 2.27 \\
		2  & 1 & 30 & $-$0.4 & $-$2.18 & 7.8  \\
		3  & 2 & 32.1 & $-$0.27 & $-$1.05 & 13.3  \\
		4  & 10 & 41.9 & $-$0.1 & $-$0.19 & 47.1  \\
		5  & 15 & 46.2 & $-$0.08 & $-$0.12 & 64.9  \\
		6  & 20 & 50 & $-$0.07 & $-$0.09 & 81.5  \\
		\bottomrule
	\end{tabular}
\caption{The values of $H_{max}$ entropy differences $\tilde{S}_S-\tilde{S}_N$ and heat capacities
$\tilde{C}_S-\tilde{C}_V$ in superconducting and normal states determined by relations \eqref{60} and \eqref{63} are presented for transition temperatures $\tilde{T}_{C_i}$, for the same values of $\tilde{\omega}_{Hcr,i}$ as in Figure~\ref{fig2}.
\label{table2}}
\end{table}

It follows from what has been said that Ginzburg-Landau temperature expansion for a critical
field near the critical temperature $T_c$ is not applicable for Bose-condensate of TI-bipolarons.
Since the temperature dependence $H_{cr}(T)$ determines the temperature dependencies
of all thermodynamic quantities, this conclusion is valid for all such values.
As was pointed out in the Introduction, this conclusion follows from the fact that BCS theory,
in view of its nonanalyticity on coupling constant, on no condition passes on to the theory of bipolaron condensate.

Above we dealt with an isotropic case. In the anisotropic case formulae \eqref{27a}, \eqref{27b} yield:

\begin{equation}\label{64}
		H^2_{max}=H^2_{max\;\bot}=\frac{2\omega _0M_{\bot}\hbar^2}{\eta^2},\ \ \ \vec{B}||\vec{c},
\end{equation}
i.e., in the case when the magnetic field is directed perpendicular to the plane of layers and:

\begin{equation}\label{65}
		H^2_{max}=H^2_{max\;||}=\frac{2\omega _0M_{||}\hbar^2}{\eta^2},\ \ \ \vec{B}\bot\vec{c},
\end{equation}
in the case when the magnetic field lies in the plane of layers. From \eqref{64} and \eqref{65} it follows that:

\begin{equation}\label{66}
		\frac{H^2_{max\;\bot}}{H^2_{max\;||}}=\sqrt{\frac{M_{\bot}}{M_{||}}}=\gamma.
\end{equation}

With the use of \eqref{54}, \eqref{65}, \eqref{66}, the critical field $H_{cr}(\tilde{T})$
(in the directions perpendicular and parallel to the plane of layers) will be:

\begin{equation}\label{67}
		 H_{cr\;||,\bot}(\tilde{T})=H_{max\;||,\bot}\sqrt{1-\tilde{\omega}_{H_{cr}}(\tilde{T})/\tilde{\omega}}.
\end{equation}

From \eqref{67} it follows that the relations $H_{cr\;||}(\tilde{T})/H_{cr\;\bot}(\tilde{T})$
are independent of temperature. The dependencies obtained are compared with experimental data in Section~\ref{Comparison}.

\section{Comparison with the Experiment}\label{Comparison}

By way of example let us consider HTSC YBa$_2$Cu$_3$O$_7$ with the temperature of transition $90\div93$ K, volume of the unit cell $0.1734\times10^{-21}$ cm$^3$, concentration of holes $n\cong10^{21}$ cm$^{-3}$.
According to estimates~\cite{56}, Fermi energy is equal to: $\varepsilon _F=0.37$ eV.
Concentration of TI-bipolarons in YBa$_2$Cu$_3$O$_7$ is found from equation \eqref{29}:
\begin{equation}\label{68}
		\frac{n_{bp}}{n}C_{bp}=f_{\tilde{\omega}}(\tilde{T}_c),
\end{equation}
with $\tilde{T}_c=1.6$.

Among experiments with the use of an external magnetic field, of importance are experiments
concerned with measurements of London penetration depth $\lambda$.
In YBa$_2$Cu$_3$O$_7$ for $\lambda$ for $T=0$ the authors of~\cite{61} obtained
$\lambda_{ab}=150\div300$ nm, $\lambda _c=800$ nm.
The same order of magnitude of these quantities is given in a lot of papers~\cite{62,63,64,65}.
The authors of~\cite{64} (see also references therein) demonstrate that anisotropy of lengths
$\lambda_a$ and $\lambda_b$ in cuprate planes can be $30\%$ depending on the type of the crystal structure.
If we take the value $\lambda _{a}=150$ nm and $\lambda _c=800$ nm obtained on most papers, then,
according to \eqref{47} the anisotropy parameter will be $\gamma\approx30$,
which is the value usually used for for YBa$_2$Cu$_3$O$_7$ crystals.

The temperature dependence $\lambda^2(0)/\lambda^2(T)$ was studied in many papers
(see~\cite{65} and references therein).

\begin{figure}
\begin{center}
\includegraphics[width=0.8\linewidth]{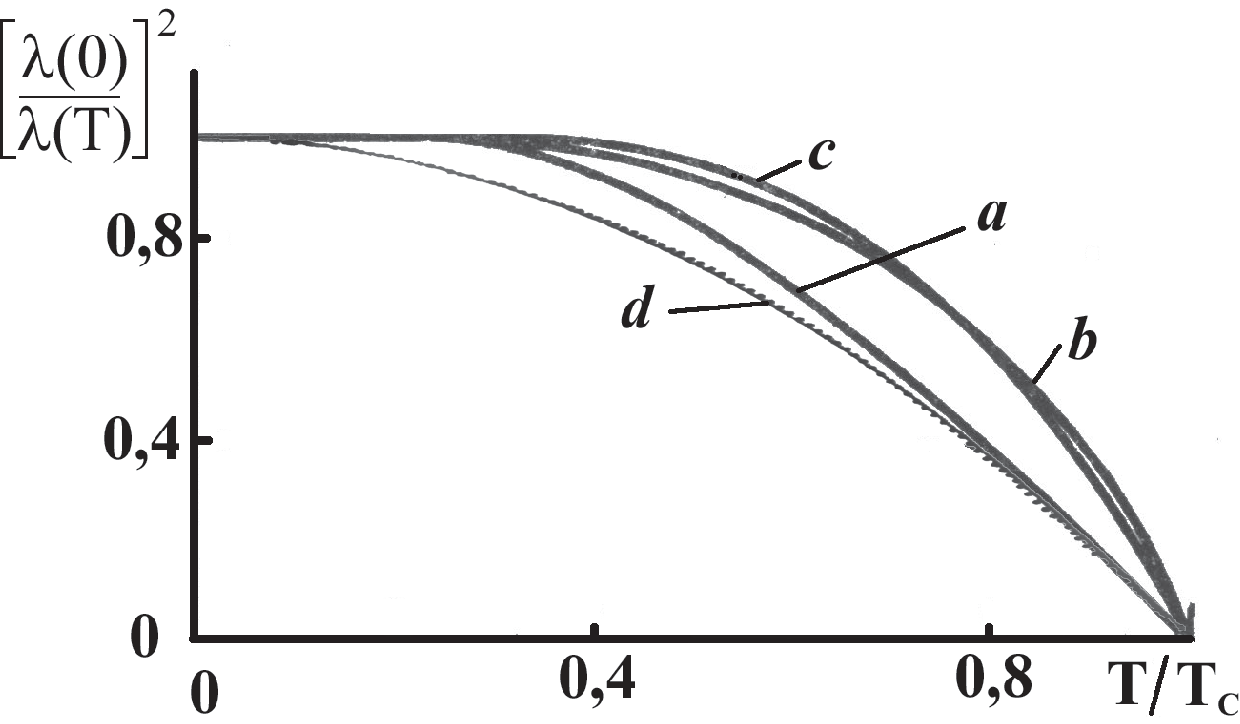}
\caption{Penetration depth of the magnetic field found with the use of BCS theory (\textsl{a}-local approximation,
\textsl{b}-nonlocal approximation); on empirical law $\lambda^{-2}~1-(T/T_c)^4$ (\textsl{c})~\cite{65a};
in YBa$_2$Cu$_3$O$_7$ (\textsl{d})~\cite{65}.}\label{fig10}
\end{center}
\end{figure}

Figure~\ref{fig10} shows a comparison of various curves for $\lambda^2(0)/\lambda^2(T)$.
In paper~\cite{65} it is shown that in high quality crystals of YBa$_2$Cu$_3$O$_7$ the temperature dependence
$\lambda^2(0)/\lambda^2(T)$ is well approximated by a simple dependence  $1-t^2$, $t=T/T_c$.

\begin{figure}
\begin{center}
\includegraphics[width=0.8\linewidth]{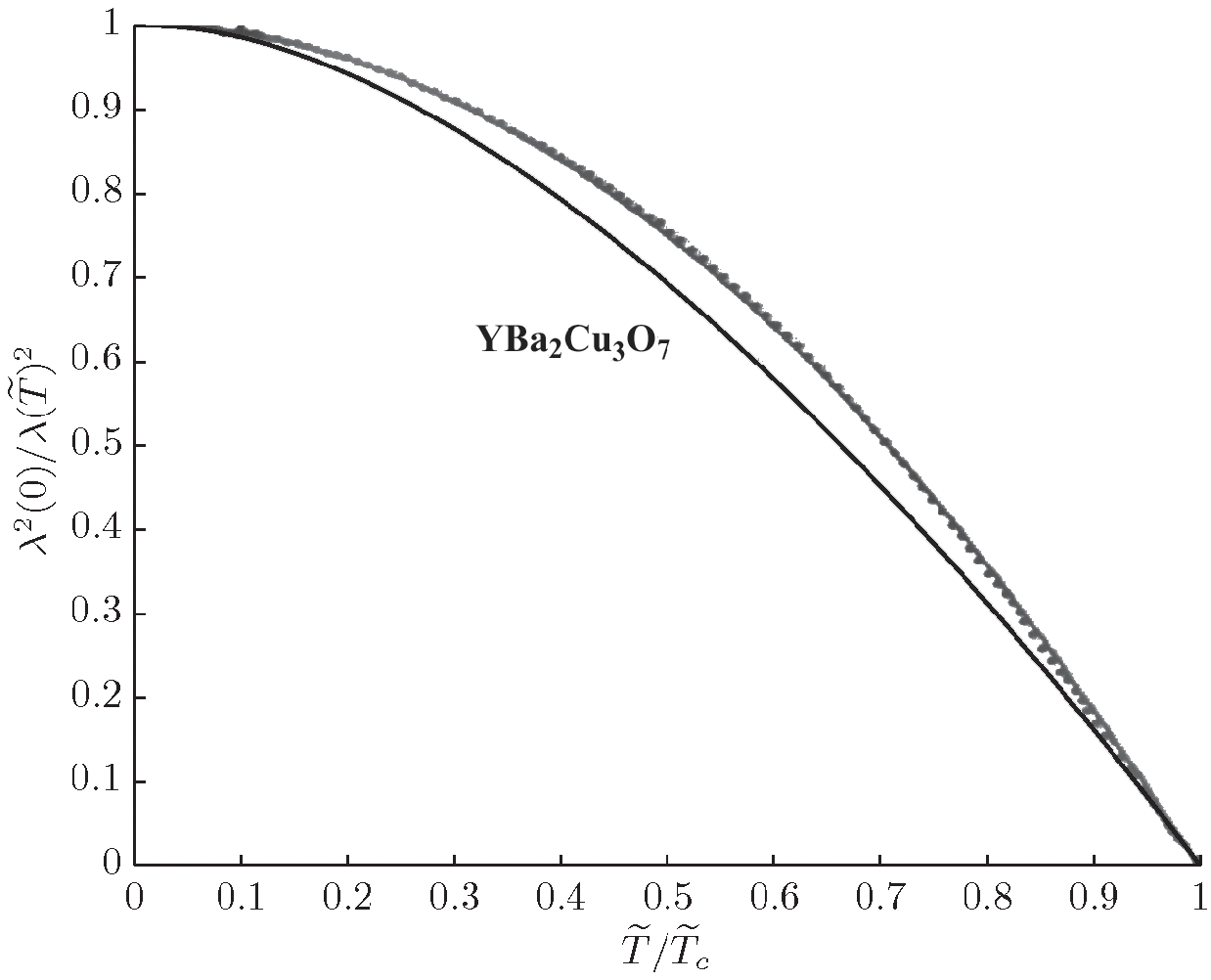}
\caption{Comparison of the theoretical dependence $\lambda^2(0)/\lambda^2(\tilde{T})$
(solid line) obtained in the present article with the experimental one~\cite{65} (dotted line).}\label{fig11}
\end{center}
\end{figure}

Figure~\ref{fig11} demonstrates a comparison of the experimental dependence $\lambda^2(0)/\lambda^2(T)$~\cite{65} with the theoretical one:

\begin{equation}\label{73}
		\frac{\lambda^2(0)}{\lambda^2(T)}=1-\left(\frac{T}{T_c}\right)^{3/2}
		\frac{F_{3/2}(\omega/T)}{F_{3/2}(\omega/T_c)},
\end{equation}
which follows from \eqref{48}, \eqref{30}.
Hence there is a good agreement between experimental and theoretical dependencies \eqref{73}.

\begin{figure}
\begin{center}
\includegraphics[width=0.8\linewidth]{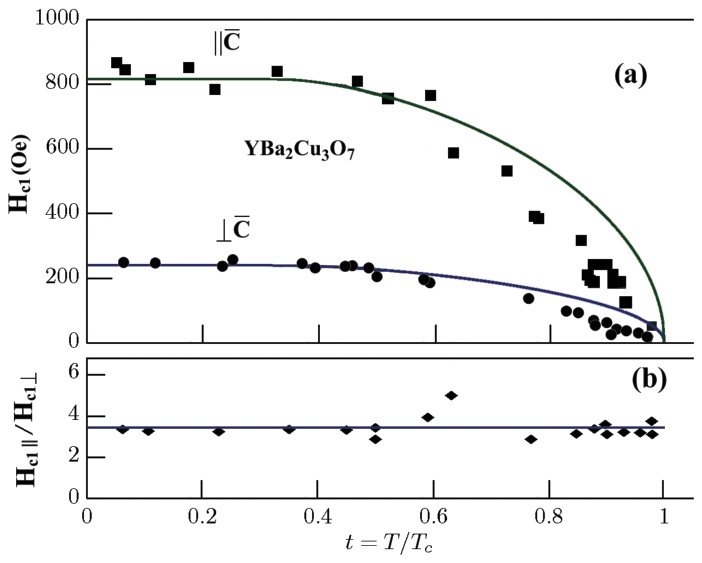}
\caption{Comparison of calculated (solid line) and experimental values of $H_{c1}$
(squares, circles, rhombs~\cite{66}) for the cases $||\vec{c}$ and $\bot\vec{c}$.}\label{fig12}
\end{center}
\end{figure}

The theory developed enables us to compare the temperature dependence of the value of
the critical magnetic field in YBa$_2$Cu$_3$O$_7$ with experimental data~\cite{66}.
Since the theory constructed in Section~\ref{Thermodyn} describes a homogeneous state of a TI-bipolaron gas,
then the critical field under consideration corresponds to a homogeneous Meissner phase.
In paper~\cite{66} this field is denoted by $H_{c1}$ which is related to denotations of Section~\ref{Thermodyn} as:
$H_{c1}=H_{cr}$, $H_{c1||}=H_{cr\bot}$, $H_{c1\bot}=H_{cr||}$.
To make a comparison with the experiment we use parameter values obtained earlier for YBa$_2$Cu$_3$O$_7$:
 $\tilde{\omega}=1.5$, \linebreak $\tilde{\omega}_c=1.6$.
Figure~\ref{fig12} shows a comparison of experimental dependencies $H_{c1\bot}(T)$ and $H_{c1||}(T)$~\cite{66}
with theoretical dependencies \eqref{67}, where for $H_{max\;||,\bot}(T)$,
we took the following experimental values: \linebreak $H_{max\;||}=240$, $H_{max\;\bot}=816$.
The results presented in Figure~\ref{fig12} confirm the conclusion (Section~\ref{Thermodyn}) that relations $H_{cr\bot}(T)/H_{cr||}(T)$ are independent of temperature.

Relations \eqref{47}, \eqref{64}, \eqref{65} yield:

\begin{equation}\label{74}
		 \left(\gamma^*\right)^2=\frac{M^*_{\bot}}{M^*_{||}}\propto\frac{\lambda^2_{\bot}}{\lambda^2_{||}};\ \ \
		\frac{H^2_{max\;\bot}}{H^2_{max\;||}}=\gamma^2=11.6.
\end{equation}

The assessment of anisotropy parameters $\gamma^2=11.6$ determined by relations \eqref{74}
differs from the value $\left(\gamma^*\right)^2=30$ used above.
This difference is probably caused by difference in anisotropy of polaron effective mass $M^*_{||,\bot}$ and electron band mass $m^*_{||,\bot}$.

\section{Scaling Relations}\label{Scaling}

Scaling relations play an important role in the theory of superconductivity assisting
the search for new high-temperature superconductors with record parameters.
These relations can emerge as a result of numerous experiments lacking any reliable
theoretical substantiation. Or else they can be derived from insufficiently reliable
theoretical considerations, but subsequently be supported by a lot of experiments.
By way of example we refer to Uemura law considered in the previous section.

The theory presented here enables us to give a natural explanation to some important scaling relations.
In particular, in this section we will derive Alexandrov's formula~\cite{67,68} and Homes's scaling~law.
\begin{itemize}
\item	{\bf Alexandrov's formula}
As was mentioned in~\cite{Lakhno-ACMP}, in an anisotropic case formula \eqref{68} takes on the form:

\begin{equation}\label{75}
		 \tilde{T}_c=F^{-2/3}_{3/2}(\tilde{\omega}/\tilde{T}_c)\left(\frac{n_{bp}}{M_{||}}\right)
		\frac{2\pi\hbar^2}{M^{1/3}_{\bot}\omega^*}.
\end{equation}
It is convenient to pass on in formula \eqref{75} from quantities $n_{bp}$, $M_{||}$, $M_{\bot}$
which can hardly be determined in experiments to quantities which are easily measured experimentally:

\begin{equation}\label{76}
		\lambda _{ab}=\left[\frac{M_{||}}{16\pi n_{bp}e^2}\right]^{1/2},\ \ \
		\lambda _{c}=\left[\frac{M_{\bot}}{16\pi n_{bp}e^2}\right]^{1/2},\ \ \
		R_H=\frac{1}{2en_{bp}},
\end{equation}
where $\lambda _{ab}=\lambda_{||}$, $\lambda_c=\lambda _{\bot}$ are London lengths
of penetration into the planes of layers and in perpendicular direction, respectively;
$R_H$ is Hall coefficient. In expressions \eqref{76} the light velocity is assumed to be equal to unity: $c=1$.
With the use of relations \eqref{76} and \eqref{75} we get:

\begin{equation}\label{77}
		k_BT_c=\frac{2^{1/3}}{8}F^{-2/3}_{3/2}\left(\tilde{\omega}/\tilde{T}_c\right)
		\frac{\hbar^2}{e^2}\left(\frac{eR_H}{\lambda^4_{ab}\lambda^2_c}\right)^{1/3}.
\end{equation}
In formula \eqref{77} the quantity $eR_H$ is measured in cm$^3$, $\lambda_{ab}$, $\lambda_c$ in cm, $T_c$ in Kelvin.

Taking into account that in most HTSC materials $\tilde{\omega}\approx\tilde{T}_c$ and the function
$F_{3/2}(\tilde{\omega}/\tilde{T})$ changes near $\tilde{\omega}=\tilde{T}_c$
only slightly, with the use of the value $F_{3/2}(1)=0.428$ and expression \eqref{76} we present $T_c$ in the form:

\begin{equation}\label{78}
		T_c\cong 8.7\left(\frac{eR_H}{\lambda^4_{ab}\lambda^2_c}\right)^{1/3}.
\end{equation}
Formula \eqref{78} differs from Alexandrov's formula~\cite{67,68} only in numerical coefficient which in~\cite{67,68} is equal to 1.64. As is shown in \cite{67,68}, formula \eqref{78} practically always properly
describes relation between the parameters for all known HTSC materials.
In~\cite{67,68} it is also shown that Uemura relation~\cite{69,69a} is a particular case of formula \eqref{78}.
\vspace{6pt}
\item	{\bf Homes's law}
Homes's law holds that scaling relation are valid for superconducting \mbox{materials~\cite{70,71}}:

\begin{equation}\label{79}
		\rho _S=C\sigma _{DC}(T_c)T_c,
\end{equation}
where $\rho _S$ is the density of a superfluid component for $T=0$, $\sigma _{DC}(T_c)$
 is the conductivity of direct current for $T=T_c$, $C$ is a constant equal to $\approx 35$cm$^{-2}$
for ordinary superconductors and HTSC materials for a current running in the plane of layers.

The quantity $\rho _S$ involved in \eqref{79} is related to plasma frequency $\omega_p=\sqrt{4\pi n_Se^2_S/m^*_S}$
(where $n_S$ is a concentration of superconducting charge carriers; $m^*_S$, $e_S$ are a mass and charge of
superconducting charge carriers) by a well-known expression $\rho _S=\omega^2_p$~\cite{72}.
Using this expression, relation $\sigma_{DC}=e^2_nn_n\tau/m^*_n$ (where $n_n$ is the concentration of charge carriers for
$T=T_c$), $m^*_n$, $e_n$ are the mass and the charge of charge carriers, relation $\tau\sim\hbar /T_c$
(where $\tau$ is the minimum Planck time for scattering of electrons at the critical point~\cite{72})
and also expression \eqref{79}, on the assumption $e_S=e_n$, $m_S=m_n$, we get:

\begin{equation}\label{80}
		n_S(0)\cong n_n(T_c).
\end{equation}

This result is confirmed by experimental data \cite{Balakirev-2003}.

In our scenario of Bose-condensation of TI-bipolarons, Homes's law in the form of \eqref{80} becomes almost obvious.
Indeed, for $T=T_c$ TI-bipolarons are stable formations (they decay at temperature equal to the pseudogap energy which considerably exceeds $T_c$). Their concentration for \mbox{$T=T_c$} is equal to $n_n$ and therefore these bipolarons for $T=T_c$
start forming condensate whose concentration $n_S(T)$ reaches its maximum $n_S(0)=n_n(T_c)$ for $T=0$
(i.e., when bipolarons fully pass on to condensed state) which corresponds to relation \eqref{80}.
Notice that in the framework of BCS theory Homes's law can hardly be explained.
\end{itemize}

\section{Summary}\label{Summary}

It is generally accepted that super-flow, superfluidity and superconductivity are collective phenomena that are driven by inter-particle interactions. Here we state an opposite suggestion that the above phenomena are mainly determined by the specific properties of separate boson particles.

In this paper we have presented conclusions emerging from consistent translation-invariant consideration of EPI.
It implies that pairing of electrons, for any coupling constant, leads to a concept
of TI-polarons and TI-bipolarons. Being bosons, TI-bipolarons can experience
Bose condensation when $H \neq 0$ leading to superconductivity.
Let us list the main results following from this approach.

First and foremost the theory resolves the problem of the great value of
the bipolaron effective mass (Section~\ref{Stat}). As a consequence, formal limitations on the value
of the critical temperature of the transition are eliminated too.
The theory quantitatively explains such thermodynamic properties of HTSC-conductors
as availability and value of the jump in the heat capacity (details in~\cite{Lakhno-ACMP}) lacking
in the theory of Bose condensation of an ideal gas. The theory also gives
an insight into the occurrence of a great ratio between the width of the pseudogap and $T_c$ (\cite{Lakhno-ACMP,lakhno-physC}).
It accounts for the small value of the correlation length~\cite{36} and explains
the availability of a gap and a pseudogap (\cite{Lakhno-ACMP,lakhno-physC}) in HTSC materials.
Accordingly, isotopic effect automatically follows from expression \eqref{30}
where the phonon frequency $\omega_0$ acts as a gap. The conclusion of the dependence of
the temperature of the transition $T_c$ on the relation ${n_{bp}}/{M_{||}}$ (see ~\cite{Lakhno-ACMP})
correlates with Alexandrov-Uemura law (Section~\ref{Scaling})
universal for all HTSC materials. It is shown that
Homes's scaling law is a natural consequence of the theory presented (Section~\ref{Scaling}).
The theory explains a wide variety of phenomena observed in a magnetic field (Section~\ref{Thermodyn}).
In particular:

\begin{enumerate}
\item	It is shown that the occurrence of a gap in the spectrum of TI-bipolarons makes possible their condensation in a magnetic field.

\item	It is demonstrated that there exists a critical value of the magnetic field above which homogeneous Bose-condensation becomes impossible.

\item	The temperature dependence of the critical magnetic field and London penetration depth obtained in the paper are in good agreement with the experiment.

\end{enumerate}

At the same time the theory presented shows that:
\begin{enumerate}
\item	Rutgers formula cannot be applied to describe Bose-condensation of TI-bipolarons.
\item	Ginzburg-Landau expansions do not suit to describe Bose-condensation of TI-bipolarons.
\end{enumerate}

The theory predict some phenomena such as:
\begin{enumerate}
\item	Isotopic effect for a jump of heat capacity in passing from the normal phase to superconducting one.
\item	A possibility of the existence of a phase transition of infinite kind in a magnetic field at low temperatures.
\item	Identity of the energy gap with phonon frequency.
\item	Existence of superconducting TI-bipolarons whose concentration is much less than the total concentration of charge carriers.
\end{enumerate}

Application of the theory to 1D and 2D systems leads to qualitatively new results since the occurrence of a gap in the TI-bipolaron spectrum automatically removes divergences at small momenta, inherent in the theory of ideal Bose gas. An important consequence of this fact is the existence of a superconducting phase in homogeneous 1D and 2D systems \cite{Lakhno-SpringerPlus}.

In conclusion it should be said that in the theory developed the TI-bipolarons are equivalent of Cooper pairing in BCS. In distinction from the latter, this theory does not impose the upper limit of critical temperature for SC transition. Of course there are a lot of other mechanisms of pairing. Each theory of superconductivity needs to explain three basic effects: the zero resistance at $T<T_c$, Meissner effect and the isotop effect. As was shown in the paper the considered EPI is sufficient for such an explanation. Recent experiments on $H_3S$ and $LaH_{10}$ (with $T_c=203$ K for $H_3S$ \cite{Drozdov-2015} and $T_c=260$ K for $LaH_{10}$ \cite{Somayazulu-2019,Drozdov-2018}) possessing a record temperature of SC transition (under high pressure) are in accordance with the idea of the important role of EPI  mechanism and TI-bipolarons considered in the paper.


\vspace{6pt}

\section*{Appendix}

Hamiltonian $H_1$ involved in \eqref{14} has the form:
\begin{equation*}\label{A1}
    H_1=\sum_{k}(V_k+f_k\hbar\omega_k)(a_k+a_k^{+})+\sum_{k,k'}\frac{\vec{k}\vec{k'}}{m}f_{k'}(a_k^{+}a_ka_{k'}+a_k^{+}a_{k'}^{+}a_k)+
    \frac{1}{2m}\sum_{k,k'}\vec{k}\vec{k'}a_k^{+}a_{k'}^{+}a_ka_{k'}\,,
\end{equation*}

Let us apply the operator $H_1$ to functional $\hat{R}|0\rangle$, where $\hat{R}$---operator that generates Bogolyubov-Tyablikov canonical transformation \eqref{17}.  We will show that $\langle0|\hat{R}^{+}H_1\hat{R}|0\rangle=0$. Indeed, the action of $\hat{R}$ on  $H_1$ terms containing
an odd number of operators in $H_1$ (i.e., the first and second terms in $H_1$) will always contain an odd number of terms and mathematical expectation for these terms will tend to zero.

Let us consider mathematical expectation for the last term in $H_1$:
\begin{equation}\label{A2}\tag {$A1$}
    \langle0|\hat{R}^{+}\sum_{k,k'}\vec{k}\vec{k'}a_k^{+}a_{k'}^{+}a_ka_{k'}\hat{R}|0\rangle\,.
\end{equation}

The function $\langle0|\hat{R}^{+}a_k^{+}a_{k'}^{+}a_ka_{k'}\hat{R}|0\rangle$ represents the norm of vector $a_ka_{k'}\hat{R}|0\rangle$
and will be positively defined for all  $k$  and $k'$. If we replace $\vec{k}\rightarrow-\vec{k}$ in \eqref{A2} than the whole expression will change the sign and, therefore, \eqref{A2} is also equal to zero. Hence $\langle0|\hat{R}^{+}H_1\hat{R}|0\rangle=0$.

As was shown in \cite{40} the explicit form for operator $\hat{R}$ is:
\begin{equation*}\label{A3}
    \hat{R}=C\exp\left\{\frac{1}{2}\sum_{k,k'}a_k^{+}A_{kk'}a_{k'}^{+}\right\}\,,
\end{equation*}
where $C$  is the normalizing constant and matrix $A$ satisfies the conditions:
\begin{equation*}\label{A4}
    A=M_2^{*}(M_1^{*})^{-1}\,,\ \ A=A^{T}\,,
\end{equation*}
where $M_1$ and $M_2$ are matrices involved in \eqref{17}.


\begin{thebibliography}{999}
\bibitem{1} Bardeen, J.; Cooper, L.N.; Schrieffer, J.R. {Theory of Superconductivity}. {\emph{Phys. Rev.}} \textbf{1957}, \emph{108}, 1175.

\bibitem{2} Schrieffer, J.R. \emph{Theory of Superconductivity}; Westview Press: Oxford, UK, 1999.

\bibitem{3} Moriya, T.; Ueda, K. Spin fluctuations and high temperature superconductivity. \emph{Adv. Phys.} \textbf{2000}, \emph{49}, 555.

\bibitem{4} Sinha, K.P.; Kakani, S.L. Fermion local charged boson model and cuprate superconductors. \emph{Proc. Natl. Acad. Sci. India Sect. A Phys. Sci.} \textbf{2002}, \emph{72}, 153.
%

\bibitem{5} Alexandrov, A.S. \emph{Theory of Superconductivity from Weak to Strong Coupling};
IoP Publishing: Bristol, UK, 2003.

\bibitem{6} Manske, D. \emph{Theory of Unconventional Superconductors}; Springer: Heidelberg, Germany, 2004.

\bibitem{7} Benneman, K.H.; Ketterson, J.B. (Eds.) \emph{Superconductivity: Conventional and
Unconventional Superconductors 1--2}; Springer: New York, NY, USA; Berlin, Germany, 2008.

\bibitem{8} Gunnarsson, O.; R\"{o}sch, O. {Interplay between electron-phonon and
coulomb interactions in cuprates}. \emph{J. Phys.} \textbf{2008}, \emph{20}, 043201.

\bibitem{9} Kakani, S.L.; Kakani, S. \emph{Superconductivity}; Anshan: Kent, UK, 2009.

\bibitem{10} Plakida, N.M. \emph{High Temperature Cuprate Superconductors: Experiment,
Theory and Applications}; Springer: Heidelberg, Germany, 2010.

\bibitem{11} Cooper, L.N.; Feldman, D. (Eds.) \emph{BCS: 50 Years}; World Sci. Publ. Co.: Singapore, 2011.

\bibitem{12} Tohyama, T. {Recent progress in physics of high-temperature superconductors}. \emph{Jpn. J. Appl. Phys.} \textbf{2012}, \emph{51}, 010004.

\bibitem{13} Askerzade, I. \emph{Unconventional Superconductors: Anisotropy and Multiband Effects}; Springer: Berlin, Germany, 2012.

\bibitem{14} Bardeen, J. {Developments of concepts in superconductivity}. \emph{Phys. Today} \textbf{1963}, \emph{16}, 19.

\bibitem{15} Keldysh, L.V.; Kozlov, A.N. {Collective Properties of Excitons in Semiconductors}. \emph{Sov. Phys. JETP} \textbf{1968}, \emph{27}, 521.

\bibitem{16} Eagles, D.M. {Possible Pairing without Superconductivity at Low Carrier Concentrations
in Bulk and Thin-Film Superconducting Semiconductors}. \emph{Phys. Rev.} \textbf{1969}, \emph{186}, 456.

\bibitem{17} Nozi\`{e}res, P.; Schmitt-Rink, S. {Propagation of Second sound in a superfluid
Fermi gas in the unitary limit}. \emph{J.~Low Temp. Phys.} \textbf{1985}, \emph{59}, 195.

\bibitem{18} Loktev, V.M. {Mechanisms of high-temperature superconductivity of Copper oxides}. \emph{Fizika Nizkih Temperatur} \textbf{1996}, \emph{22}, 3.

\bibitem{19} Randeria, M. {Precursor Pairing Correlations and Pseudogaps}. \emph{arXiv} \textbf{1997}, arXiv:cond-mat/9710223.

\bibitem{20} Uemura, Y.J. {Bose-Einstein to BCS crossover picture for high-$T_c$ cuprates}.
\emph{Phys. C Supercond.} \textbf{1997}, \emph{282}, 194--197.

\bibitem{21} Drechsler, M.; Zwerger, W. {Crossover from BCS-superconductivity to Bose-condensation}. \emph{Ann. Phys.} \textbf{1992}, \emph{1}, 15.

\bibitem{22} Griffin, A.; Snoke, D.W.; Stringari, S. (Eds.)
\emph{Bose-Einstein Condensation}; Cambridge University Press: New~York, NY, USA, 1996.

\bibitem{23} Eliashberg, G.M. {Interactions between Electrons and Lattice Vibrations in a Superconductor}. \emph{Sov. Phys. JETP} \textbf{1960}, \emph{11}, 696.

\bibitem{24} Marsiglio, F.; Carbotte, J.P. {Gap function and density of states in the strong-coupling limit
for an electron-boson system}. \emph{Phys. Rev. B} \textbf{1991}, \emph{43}, 5355.

\bibitem{25} Micnas, R.; Ranninger, J.; Robaszkiewicz, S. {Superconductivity in narrow-band systems with
local nonretarded attractive interactions}. \emph{Rev. Mod. Phys.} \textbf{1990}, \emph{62}, 113.

\bibitem{26} Zwerger, W. (Ed.) {The BCS-BEC Crossover and the Unitary Fermi Gas}. In \emph{Lecture Notes in Physics};
Springer: Berlin, Heidelberg, 2012.

\bibitem{27} Bloch, I.; Dalibard, J.; Zwerger, W. {Many-body physics with ultracold gases}. \emph{Rev. Mod. Phys.} \textbf{2008}, \emph{80}, 885.

\bibitem{28} Giorgini, S.; Pitaevskii, L.P.; Stringari, S. {Theory of ultracold atomic Fermi gases}.
\emph{Rev. Mod. Phys.} \textbf{2008}, \emph{80}, 1215.

\bibitem{29} Chen, Q.; Stajic, J.; Tan, S.; Levin, K. {BCS-BEC crossover: From high temperature superconductors
to ultracold superfluids}. \emph{Phys. Rep.} \textbf{2005}, \emph{412}, 1--88.

\bibitem{30} Ketterle, W.; Zwierlein, M.W. {Making, probing and understanding ultracold Fermi gases}. In \emph{Ultra-cold Fermi Gases}; Inguscio, M., Ketterle, W., Salomon, C., Eds.; IOS Press: Amsterdam, The Netherlands, 2007; p. 95.

\bibitem{31} Pieri, P.; Strinati, G.C. {Strong-coupling limit in the evolution from BCS superconductivity to Bose-Einstein condensation}. \emph{Phys. Rev. B} \textbf{2000}, \emph{61}, 15370.

\bibitem{32} Gerlach, B.; L\"{o}wen, H. {Analytical properties of polaron systems or:
Do polaronic phase transitions exist or not?} \emph{Rev. Mod. Phys.} \textbf{1991}, \emph{63}, 63.

\bibitem{33} Lakhno, V.D. Translation invariant theory of polaron (bipolaron) and the problem of quantizing near the classical solution. \emph{JETP} \textbf{2013}, \emph{116}, 892. doi:10.1134/S1063776113060083.

\bibitem{34} Gor'kov, L.P. {Microscopic derivation of the Ginzburg-Landau equations in the theory of superconductivity}. \emph{Sov. Phys. JETP} \textbf{1959}, \emph{9}, 1364.

\bibitem{35} Lakhno, V.D. Energy and Critical Ionic-Bond Parameter of a 3D Large-Radius Bipolaron. \emph{JETP} \textbf{2010}, \emph{110}, 811.

\bibitem{36} Lakhno, V.D. Translation-invariant bipolarons and the problem of high-temperature superconductivity. \emph{Sol.~State Commun.} \textbf{2012}, \emph{152}, 621.

\bibitem{37} Kashirina, N.I.; Lakhno, V.D.; Tulub, A.V. The Virial Theorem and the Ground State Problem in Polaron Theory. \emph{JETP} \textbf{2012}, \emph{114}, 867.

\bibitem{38} Lakhno, V.D. Pekar's ansatz and the strong coupling problem in polaron theory. \emph{Phys. Usp.} \textbf{2015}, \emph{58}, 295.

\bibitem{Lakhno-ACMP} Lakhno, V.D. Superconducting Properties of 3D Low-Density Translation-Invariant Bipolaron Gas. \emph{Adv.~Condens. Matter Phys.} \textbf{2018}, \emph{2018}, 1380986. doi:10.1155/2018/1380986.

\bibitem{lakhno-physC} Lakhno, V.D. Superconducting properties of a nonideal bipolaron gas. \emph{Phys. C Supercond. Its Appl.} \textbf{2019}, \emph{561}, 1--8. doi:10.1016/j.physc.2018.10.009.

\bibitem{39} Lakhno, V.D. Spin wave amplification in magnetically ordered crystals. \emph{Phys. Usp.} \textbf{1996}, \emph{39}, 669.

\bibitem{40} Tulub, A.V. Slow Electrons in Polar Crystals. \emph{Sov. Phys. JETP} \textbf{1962}, \emph{14}, 1301.

\bibitem{41} Heisenberg, W. {Die selbstenergie des elektrons}. \emph{Z. Phys.} \textbf{1930}, \emph{65}, 4.

\bibitem{42} Rosenfeld, L. {\"{U}ber eine m\"{o}gliche Fassung des Diracschen Programms zur Quantenelektrodynamik
und deren formalen Zusammenhang mit der Heisenberg-Paulischen Theorie}. \emph{Z. Phys.} \textbf{1932}, \emph{76}, 729.

\bibitem{43} Lee, T.D.; Low, F.; Pines, D. {The motion of electrons in a polar crystal}.
\emph{Phys. Rev.} \textbf{1953}, \emph{90}, 297.

\bibitem{44} Tyablikov, S.V. \emph{Methods in the Quantum Theory of Magnetism}; Plenum Press: New York, NY, USA, 1967.

\bibitem{44-Miyake} Miyake, S.J. {Bound Polaron in the Strong-coupling Regime}. In
\emph{Polarons and Applications}; Lakhno, V.D., Ed.; Wiley: Leeds, UK, 1994; p. 219.

\bibitem{45-Levenson} Levinson, I.B.; Rashba, E.I. Threshold phenomena and bound states in the polaron problem. \emph{Sov. Phys. Usp.} \textbf{1974}, \emph{16}, 892--912.

\bibitem{45} Porsch, M.; R\"{o}seler, J. {Recoil Effects in the Polaron Problem}.
\emph{Phys. Status Solidi B} \textbf{1967}, \emph{23}, 365.

\bibitem{46} Alexandrov, A.S.; Mott, N. \emph{Polarons \& Bipolarons}; World Sci. Pub. Co.: Singapore, 1996.

\bibitem{47} Alexandrov, A.S.; Krebs, A.B. Polarons in high-temperature superconductors. \emph{Sov. Phys. Usp.} \textbf{1992}, \emph{35}, 345, 383. doi:10.1070/PU1992v035n05ABEH002235.

\bibitem{48} Ogg, R.A., Jr. {Superconductivity in solid metal-ammonia solutions}.
\emph{Phys. Rev.} \textbf{1946}, \emph{70}, 93.

\bibitem{49} Vinetskii, V.L.; Pashitskii, E.A. Superfluidity of charged Bose-gas and bipolaron mechanism of superconductivity. \emph{Ukr. J. Phys.} \textbf{1975}, \emph{20}, 338.

\bibitem{50} Pashitskii, E.A.; Vinetskii, V.L. Plasmon and bipolaron mechanisms of high-temperature superconductivity. \emph{JETP Lett.} \textbf{1987}, \emph{46}, S104.

\bibitem{51} Emin, D. {Formation, motion, and high-temperature superconductivity of large bipolarons}. \emph{Phys. Rev. Lett.} \textbf{1989}, \emph{62}, 1544.

\bibitem{52} Vinetskii, V.L.; Kashirina, N.I.; Pashitskii, E.A. Bipolaron states in ion crystals and the problem of high temperature superconductivity. \emph{Ukr. J. Phys.} \textbf{1992}, \emph{37}, 76.

\bibitem{52a} Emin, D. {In-plane conductivity of a layered large-bipolaron liquid}. \emph{Philos. Mag.} \textbf{2015}, \emph{95}, 918.

\bibitem{53} Schmidt, V.V. \emph{The Physics of Superconductors}; Muller, P., Ustinov, A.V., Eds.; Springer: Berlin/Heidelberg, Germnay, 1997.

\bibitem{54} Pippard, A.B. {Field variation of the superconducting penetration depth}. \emph{Proc. Roy. Soc. (Lond.)}
\textbf{1950}, \emph{A203}, 210.

\bibitem{55} Schafroth, M.R. {Superconductivity of a Charged Ideal Bose Gas}. \emph{Phys. Rev.} \textbf{1955}, \emph{100}, 463.

\bibitem{56} Gor'kov, L.P.; Kopnin, N.B. High-$T_c$ superconductors from the experimental point of view. \emph{Sov. Phys. Usp.} \textbf{1988}, \emph{31}, 850.





\bibitem{61} Buckel, W.; Kleiner, R. \emph{Superconductivity: Fundamentals and Applications}, 2nd ed.; Wiley-VCH: Weinheim, Germany, 2004.

\bibitem{62} Edstam, J.; Olsson, H.K. {London penetration depth of YBCO in the frequency range 80-700 GHz}.
\emph{Physica B} \textbf{1994}, \emph{194--196 Pt 2}, 1589--1590.

\bibitem{63} Panagopoulos, C.; Cooper, J.R.; Xiang, T. {Systematic behavior of the in-plane penetration depth in d-wave cuprates}.
\emph{Phys. Rev. B} \textbf{1998}, \emph{57}, 13422.

\bibitem{64} Pereg-Barnea, T.; Turner, P.J.; Harris, R.; Mullins, G.K.; Bobowski, J.S.; Raudsepp, M.; Liang, R.; Bonn, D.A.; Hardy, W.N. {Absolute values of the London penetration depth in
YBa$_2$Cu$_3$O$_{6+y}$ measured by zero field ESR spectroscopy on Gd doped single crystals}. \emph{Phys. Rev. B} \textbf{2004}, \emph{69}, 184513.

\bibitem{65} Bonn, D.A.; Liang, R.; Riseman, T.M.; Baar, D.J.; Morgan, D.C.; Zhang, K.; Dosanjh, P.; Duty, T.L.; MacFarlane,~A.; Morris, G.D.; et al.
{Microwave determination of the quasiparticle scattering time in YBa$_2$Cu$_3$O$_{6.95}$}.
\emph{Phys. Rev. B} \textbf{1993}, \emph{47}, 11314.

\bibitem{65a} Madelung, O. \emph{Festk\"{o}rpertheorie I, II}; Springer: Berlin/Heidelberg, Germany; New York, NY, USA, 1972.

\bibitem{66} Wu, D.H.; Sridhar, S. {Pinning forces and lower critical fields in $YBa_2Cu_3O_y$ crystals:
Temperature dependence and anisotropy}. \emph{Phys. Rev. Lett.} \textbf{1990}, \emph{65}, 2074.

\bibitem{67} Alexandrov, A.S. {Comment on Experimental and Theoretical Constraints of Bipolaronic
Superconductivity in High $T_c$ Materials: An Impossibility}. \emph{Phys. Rev. Lett.} \textbf{1999}, \emph{82}, 2620.

\bibitem{68} Alexandrov, A.S.; Kabanov, V.V. {Parameter-free expression for superconducting $T_c$ in cuprates}.
\emph{Phys. Rev. B} \textbf{1999}, \emph{59}, 13628.

\bibitem{69} Uemura, Y.J.; Luke, G.M.; Sternlieb, B.J.; Brewer, J.H.; Carolan, J.F.; Hardy, W.N.; Kadono, R.; Kempton, J.R.; Kiefl, R.F.; Kreitzman, S.R.; et al. {Universal correlations between $T_c$ and ns/m (carrier density over effective mass) in high-$T_c$ cuprate superconductors.}
\emph{Phys. Rev. Lett.} \textbf{1989}, \emph{62}, 2317.

\bibitem{69a} Uemura, Y.J.; Le, L.P.; Luke, G.M.; Sternlieb, B.J.; Wu, W.D.; Brewer, J.H.; Riseman, T.M.; Seaman, C.L.; Maple, M.B.; Ishikawa, M.; et al. {Basic similarities among cuprate, bismuthate, organic,
Chevrel-phase, and heavy-fermion superconductors shown by penetration-depth measurements}. \emph{Phys. Rev. Lett.} \textbf{1991}, \emph{66}, 2665.

\bibitem{70} Homes, C.C.; Dordevic, S.V.; Strongin, M.; Bonn, D.A.; Liang, R.; Hardy, W.N.; Komiya, S.; Ando, Y.; Yu, G.; Kaneko, N.; et al. {A universal scaling relation in
high-temperature superconductors}. \emph{Nature} \textbf{2004}, \emph{430}, 539.

\bibitem{71} Zaanen, J. {Superconductivity: Why the temperature is high}. \emph{Nature} \textbf{2004}, \emph{430}, 512.

\bibitem{72} Erdmenger, J.; Kerner, P.; M\"{u}ller, S. {Towards a holographic realization of Homes law}. \emph{J. High Energy Phys.} \textbf{2012}, \emph{10}, 21.

\bibitem{Balakirev-2003} Balakirev, F.F.; Betts, J.B.; Migliori, A.; Ono, S.; Ando, Y.; Boebinger, G.S. Signature of optimal doping in Hall-effect measurements on a high-temperature superconductor. \emph{Nature} \textbf{2003}, \emph{424}, 912--915.



\bibitem{Lakhno-SpringerPlus} Lakhno, V.D. {A Translation invariant bipolaron in the Holstein model and superconductivity}. \emph{SpringerPlus} \textbf{2016}, \emph{5}, 1277. doi:10.1186/s40064-016-2975-x.

\bibitem{Drozdov-2015} Drozdov, A.P.; Eremets, M.I.; Troyan, I.A.; Ksenofontov, V.; Shylin, S.I. {Conventional superconductivity at 203 kelvin at high pressures in the sulfur hydride system}. \emph{Nature} \textbf{2015}, \emph{525}, 73--76.

\bibitem{Somayazulu-2019} Somayazulu, M.; Ahart, M.; Mishra, A.K.; Geballe, Z.M.; Baldini, M.; Meng, Y.; Struzhkin, V.V.; Hemley, R.J. Evidence for Superconductivity above 260 K in Lanthanum Superhydride at Megabar Pressures. \emph{Phys. Rev. Lett.} \textbf{2019}, \emph{122}, 027001. doi:10.1103/PhysRevLett.122.027001.

\bibitem{Drozdov-2018} Drozdov, A.P.; Kong, P.P.; Minkov, V.S.; Besedin, S.P.; Kuzovnikov, M.A.; Mozaffari, S.; Balicas, L.; Balakirev,~F.; Graf, D.; Prakapenka, V.B.; et al. Superconductivity at 250 K in lanthanum hydride under high pressures. \emph{arXiv} \textbf{2018}, arXiv:1812.01561.
\end{thebibliography}
\end{document}